\documentclass[onecolumn]{aastex631}

\usepackage{textcomp}
\usepackage{graphicx}
\usepackage{amssymb}
\usepackage{amsmath}
\usepackage{epstopdf}
\usepackage{gensymb}
\usepackage{natbib}
\usepackage{tabu}
\usepackage{color}
\usepackage{comment}
\usepackage{multirow}
\usepackage{booktabs}
\usepackage{ulem}
\usepackage{threeparttable}

\begin{document}
\title{Decoding the Radial Velocity Signatures of Solar Faculae with 3D MHD Simulations}
\shortauthors{Kr\"oll et al.}

\correspondingauthor{F.~Kr\"oll}
\email{florian.kroell@uni-graz.at}

\author[0009-0005-9418-4254]{F.~Kr\"oll}
\affiliation{Institute of Physics, University of Graz, Universit\"atsplatz 5, 8010 Graz, Austria}

\author[0000-0002-3243-1230]{K.~Sowmya}
\affiliation{Institute of Physics, University of Graz, Universit\"atsplatz 5, 8010 Graz, Austria}

\author[0000-0002-8842-5403]{A.~I.~Shapiro}
\affiliation{Institute of Physics, University of Graz, Universit\"atsplatz 5, 8010 Graz, Austria}
\affiliation{Max-Planck-Institut f\"ur Sonnensystemforschung, Justus-von-Liebig-Weg 3, 37077 G\"ottingen, Germany}

\author[0000-0002-8863-7828]{A.~Collier~Cameron}
\affiliation{Centre for Exoplanet Science, SUPA School of Physics and Astronomy, University of St Andrews, North Haugh, St Andrews KY16 9SS, UK}

\author[0000-0002-0929-1612]{V.~Witzke}
\affiliation{Institute of Physics, University of Graz, Universit\"atsplatz 5, 8010 Graz, Austria}

\author[0000-0002-3418-8449]{S.~K.~Solanki}
\affiliation{Max-Planck-Institut f\"ur Sonnensystemforschung, Justus-von-Liebig-Weg 3, 37077 G\"ottingen, Germany}
\affiliation{School of Space Research, Kyung Hee University, Yongin, Gyeonggi 17104, Republic of Korea}

\author[0000-0002-6689-0312]{I.~Ribas}
\affiliation{Institut de Ciències de l’Espai (ICE, CSIC), Campus UAB, c/ de Can Magrans s/n, 08193 Cerdanyola del Vallès, Barcelona, Spain}
\affiliation{Institut d’Estudis Espacials de Catalunya (IEEC), C/ Esteve Terrades 1, Edifici RDIT, 08860 Castelldefels, Spain}

\author[0000-0002-6436-9347]{S.~Shelyag}
\affiliation{College of Science and Engineering, Flinders University, GPO Box 2100, Adelaide, SA 5001, Australia}

\author[0000-0003-1831-7789]{G.~Kopp}
\affiliation{Institute of Physics, University of Graz, Universit\"atsplatz 5, 8010 Graz, Austria}
\affiliation{Laboratory for Atmospheric and Space Physics (LASP), University of Colorado Boulder, 1234 Innovation Drive, Boulder, Colorado 80303, USA}

\author[0000-0001-6090-1247]{N.~E.~Nemec}
\affiliation{Institut de Ciències de l’Espai (ICE, CSIC), Campus UAB, c/ de Can Magrans s/n, 08193 Cerdanyola del Vallès, Barcelona, Spain}
\affiliation{Institut d’Estudis Espacials de Catalunya (IEEC), C/ Esteve Terrades 1, Edifici RDIT, 08860 Castelldefels, Spain}

\author[0009-0005-5102-6969]{S.~Stucki}
\affiliation{Institut de Ciències de l’Espai (ICE, CSIC), Campus UAB, c/ de Can Magrans s/n, 08193 Cerdanyola del Vallès, Barcelona, Spain}
\affiliation{Institut d’Estudis Espacials de Catalunya (IEEC), C/ Esteve Terrades 1, Edifici RDIT, 08860 Castelldefels, Spain}

\begin{abstract}
We model the solar radial velocity (RV) signal induced by faculae, the dominant contributor to RV variability in Sun-like stars. We use a representative case of a facular patch transiting the visible solar disk as the Sun rotates to disentangle various physical effects contributing to the RV signal. 
Our approach is based on 3D radiative magnetohydrodynamic (MHD) simulations of the solar photosphere and upper convection zone with the MURaM code and spectral synthesis with the MPS-ATLAS code. We show that the faculae-induced RV strongly depends on the facular position on the solar disk. Near disk centre, facular magnetic fields inhibit the convective blueshift and thus produce a relative redshift of the solar spectrum. Surprisingly, when located closer to the limb, namely at heliocentric angles greater than about $60^\circ$, faculae produce a relative blueshift. This transition from redshift to blueshift is caused by the effect of magnetic fields on horizontal flows, which dominate the signal near the limb, and on the visibility of these flows. In combination with solar rotation, this centre-to-limb dependence of the facular effect leads to a complex RV profile during the facular transit and, in particular, to a phase lag between the maximum of the RV signal and the facular crossing of the central meridian. We further show that, in contrast to stellar reflex motion, the facular signal strongly depends on the spectral line in which it is measured.
\end{abstract}

\section{Introduction}
The radial velocity (RV) method has historically been one of the most successful techniques for exoplanet discoveries. While the advent of large-scale transit photometry missions has shifted the primary mode of exoplanet discovery away from RV surveys, RV observations continue to play a central role in the physical characterization of exoplanets. In this context, RV measurements are primarily used to determine planetary masses and, in combination with radii, to infer bulk densities and the nature of planets. Over the past few decades, the precision of RV spectrographs has advanced substantially --- from about 70~m/s in 1995, when the first exoplanet around a main-sequence star was detected \citep{MayorandQueloz1995}, to the current frontier of $\sim$10~cm/s achieved by modern extreme-precision RV (EPRV) instruments \citep{Figueiraetal2025}. These capabilities, in principle, allow for the detection and mass measurement of rocky planets in the habitable zones of G- and K-type dwarfs. Despite this major progress, Earth-twin discoveries and characterisations have not yet been achieved and remain a significant challenge due to the $\sim$1~m/s limitation caused by intrinsic stellar variability, known as \textit{RV jitter} \citep[see Figure~1 in][]{Johnetal2023}.

This challenge stems from the way stellar RVs are measured. In the idealised case, an observer would determine a star’s RV by measuring the positions of its spectral lines and converting their Doppler shifts relative to laboratory rest wavelengths into a stellar RV.
In practice, however, spectral lines form in highly dynamic stellar atmospheres, so that the measured Doppler shifts are affected by atmospheric gas motions. As a result, the inferred RVs deviate from the true centre-of-mass velocity and exhibit time-dependent RV jitter at the meter-per-second level \citep[see, e.g., discussion in][]{Dravinsetal2023}.

The RV jitter has two main sources. First, it arises from flows in stellar atmospheres associated with granulation, supergranulation, magnetic features, and oscillations. While these flows almost cancel out in the disk-integrated spectrum, a time-dependent imbalance remains and affects positions and shapes of spectral lines \citep{Dravinsetal1981,Dumusqueetal2011}. The second source is stellar rotation coupled with the photometric effects of magnetic features, such as bright faculae/plage and dark spots \citep{Meunieretal2010,Lagrangeetal2010,SaarandDonahue1997}.
The magnetic features on a rotating star break the symmetry between approaching and receding stellar hemispheres and lead to the RV signal.
For example, a spot on the approaching hemisphere would decrease the number of photons from it and, accordingly, its contribution to the line profile \citep{SaarandDonahue1997,Quelozetal2001}. This will produce an apparent rotation-driven net RV signal.

A variety of approaches have been developed to mitigate RV jitter. One promising direction is to build on the differential sensitivity of spectral lines to magnetic activity. While stellar Keplerian reflexive motions affect all spectral lines equally, the effect of stellar activity is line-dependent.
The differential effect cannot be utilised when spectral lines are averaged into a single cross-correlation function, a procedure that was traditionally used to achieve adequate signal-to-noise ratios. Critically, advances in instrumental stability over the last decade have made it possible to either measure RVs in individual spectral lines \citep[line-by-line approach, LBL,][]{Dumusque2018} or to group (or weigh) spectral lines according to their sensitivity to magnetic activity, thereby fully exploiting their differential behaviour.

The common feature of RV-jitter mitigating approaches based on differential line sensitivity is that they all require estimates of line sensitivities to stellar activity. Until now, these estimates have either been obtained directly from observations \citep{Dumusque2018,Thompson2020,Cretignier2020}, from simplified theoretical considerations \citep{AlMoulla2022, AlMoulla2024}, or from models based on a parametrised treatment of magnetic features(e.g., SOAP~2.0, \citealp{Dumusque2014}, or StarSim, \citealp{Herreroetal2016})

These pioneering studies have significantly advanced our understanding, yet they were limited in their ability to consistently account for the complex structure of stellar near-surface magnetoconvection, whose manifestation also depends on stellar fundamental parameters. This limitation can now be overcome thanks to advances in realistic 3D radiative magnetohydrodynamic (MHD) simulations of stellar atmospheres. As a result, first estimates of line sensitivity to stellar activity are becoming available. Until now, they have primarily been aimed at quantifying spectral line sensitivities to granulation, e.g., \cite{Frame2025, Sowmyaetal2025}, who use solar simulations from \cite{Voegleretal2005, Witzke2024} computed with MURaM \citep{Voegleretal2004}; and \cite{Dravinsetal2023, Dravinsetal2024}, who use solar simulations with CO$^5$BOLD \citep[][see also \citealt{Tremblay2013} for details]{Freytag2012}.

In this paper, we further build on the realistic MHD simulations of a quiet star and faculae from MURaM and model the signatures that faculae imprint in spectral lines. Studying spectral line sensitivity to faculae is important because for G- and K-dwarfs with a near-solar level of magnetic activity, faculae are the dominating magnetic source of the RV variations \citep{Meunieretal2010,Dumusque2014,Haywood2016,Milbourne2019}. Here we start with the exemplary case of the Sun which is also of special interest because of the recent release of long time-series of Sun-as-a-star EPRV observations \citep[from HARPS-N;][see also \citealt{Dumusqueetal2021,Kleinetal2024}]{Dumusqueetal2025}.

To this end, we use the MPS-ATLAS radiative transfer code \citep{Witzke2021} to calculate high-resolution spectra emerging from 3D MURaM models of faculae and the quiet Sun (QS). Using these spectra, we model the transit of a facular patch along the solar equator as the Sun rotates. We then analyse the spectral signatures of this facular transit in a set of Fe\,I and Fe\,II lines and show that the resulting RV signals vary strongly across these lines. This motivates line-by-line (LBL) RV extraction with customised masks and optimised line weights tailored to activity sensitivity. The simple framework we use here provides a clean, yet physically grounded baseline for separating geometric, photometric, and convective contributions to faculae-driven RV signals, paving the way for more comprehensive future studies.

The paper is organised as follows. In Section~\ref{sec:Methodology}, we introduce the spectral input, the setup of the facular transit experiment, and the RV extraction methods. In Section~\ref{sec:Results}, we start by presenting the full time-dependent RV signals resulting from the facular patch transit (hereafter, RV profiles) measured in different spectral lines and then analyse the individual physical processes that shape them. In Section~\ref{sec:summary}, we summarise the main conclusions. An additional parameter study is presented in Appendix~\ref{subsec:appendix_add-scenarios}.

\section{Methods}\label{sec:Methodology}

\subsection{MURaM and MPS-ATLAS Spectral Input}\label{subsec:spectral_input}

Our spectral input consists of QS and facular intensity spectra synthesised with MPS-ATLAS from 3D radiative-MHD MURaM simulations. We use spatially and temporally averaged spectra computed at ten viewing angles ($\mu=0.1$ -- $1.0$ in steps of 0.1, where $\mu$ is the cosine of the heliocentric angle), which capture the centre-to-limb variation (CLV) of the intensity and convective Doppler shifts.
We synthesise a set of fully isolated (blend-free) Fe\,I and Fe\,II lines at a very high resolving power of 2,000,000 to quantify subtle faculae-induced changes in line shape and wavelength shift without contamination from blends. A detailed description of the simulations, calculations of synthetic spectra, and the list of spectral lines analysed in this study are provided in Appendix~\ref{subsec:appendix_spectral_input}. We remark that the Zeeman effect is not included in the spectral synthesis, because its contribution to the facula-induced RV signal is generally expected to be small for standard RV measurements (with significant contributions limited to a few highly sensitive lines). This follows the standard approach in the literature  \citep{Meunieretal2010}. The exact influence of Zeeman effect on the faculae-induced RV signal will be discussed in an upcoming study.

\begin{figure}[ht]
    \centering
    \includegraphics[width=0.7
        \linewidth]{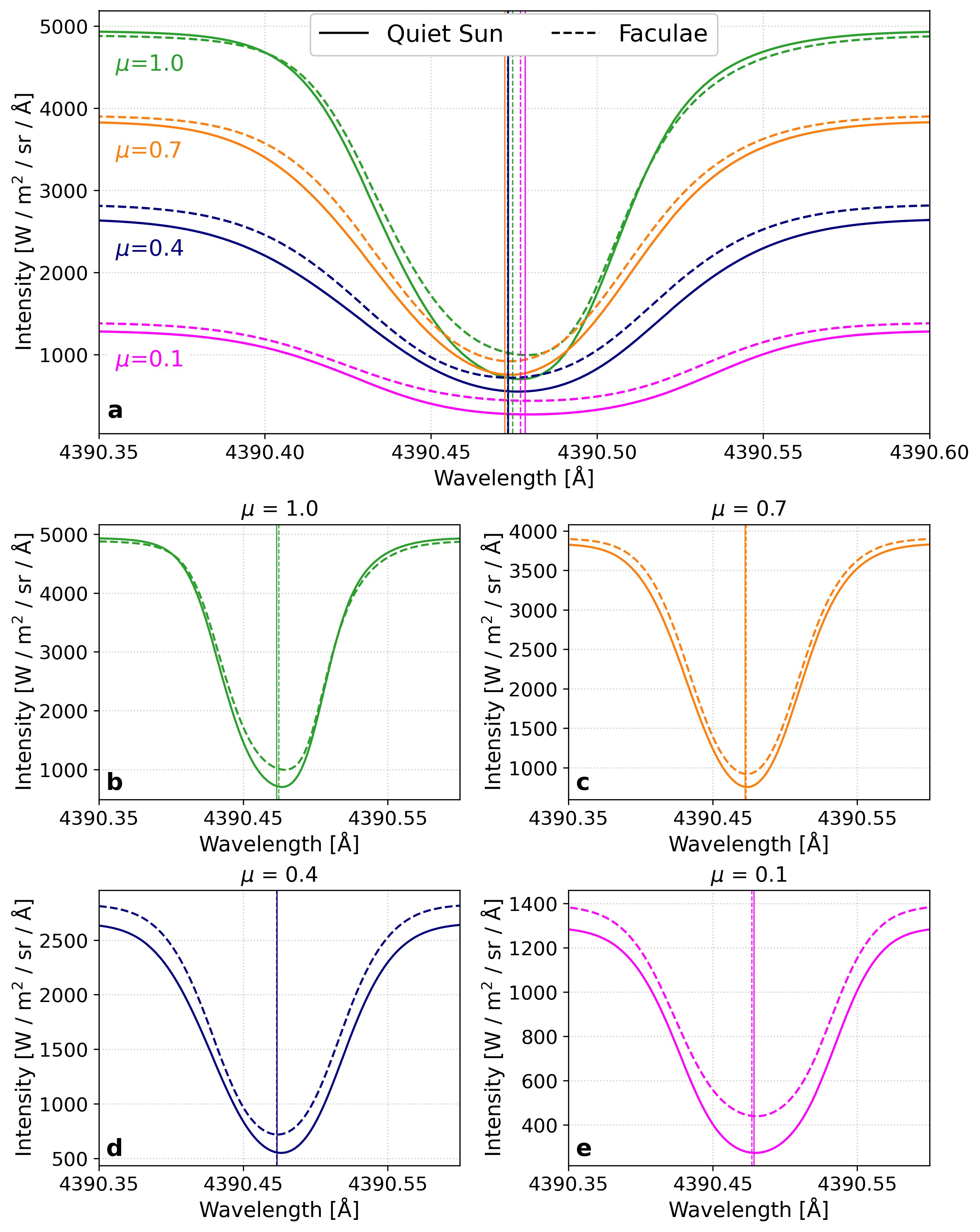}
    \caption{Comparison of the quiet Sun (QS; solid) and faculae (dashed) intensity spectra for the representative Fe\,I\,$\lambda$4390 line (Table~\ref{FeI_lines} in Appendix~\ref{subsec:appendix_spectral_input}). Panel (a): spectra at four disk positions going from disk centre ($\mu=1$) towards the limb ($\mu=0.1$). The solid and dashed vertical lines denote the position of the line centre of gravity (COG) for QS and faculae profiles, respectively. Panels (b)--(e): line profiles for individual $\mu$ as indicated in the titles. The vertical lines denote COG, as before. We note that the relative shift of the COG between QS and faculae changes sign towards smaller $\mu$.}
   \label{fig:mu_spectra}
\end{figure}

A comparison of the QS and facular spectra for the Fe\,I\,$\lambda$4390 line presented in Figure~\ref{fig:mu_spectra} highlights several key features.
First, the line profiles are always asymmetric. Second, faculae are brighter than the QS at all $\mu$ values except in the continuum at disk centre \citep[consistent with the findings from e.g.][]{Spruit1976,Topkaetal1992,Lawrenceetal1993}. Third, from disk centre towards the limb, the continuum intensities decrease more rapidly than the line-core intensities, and the intensity contrast between QS and faculae increases. Fourth, the spectral line positions exhibit a non-monotonic centre-to-limb variation (Figure~\ref{fig:mu_spectra}a). The relative line shifts can be identified from the positions of the vertical markers, with a displacement to the right indicating a redshift and one to the left a blueshift. At disk centre, the facular profile is redshifted relative to the QS profile. The magnitude of this relative redshift decreases towards the limb and eventually changes sign, such that the facular profile becomes blueshifted closer to the limb (Figure~\ref{fig:mu_spectra}b--e). We discuss the physical origin of these characteristic features and how they translate into faculae-induced RV signals further in Section~\ref{sec:Results}.

\subsection{Modelling the facular transit}\label{subsec:sim_setup}

\begin{figure}[ht]
    \centering
    \includegraphics[width=0.6
        \linewidth]{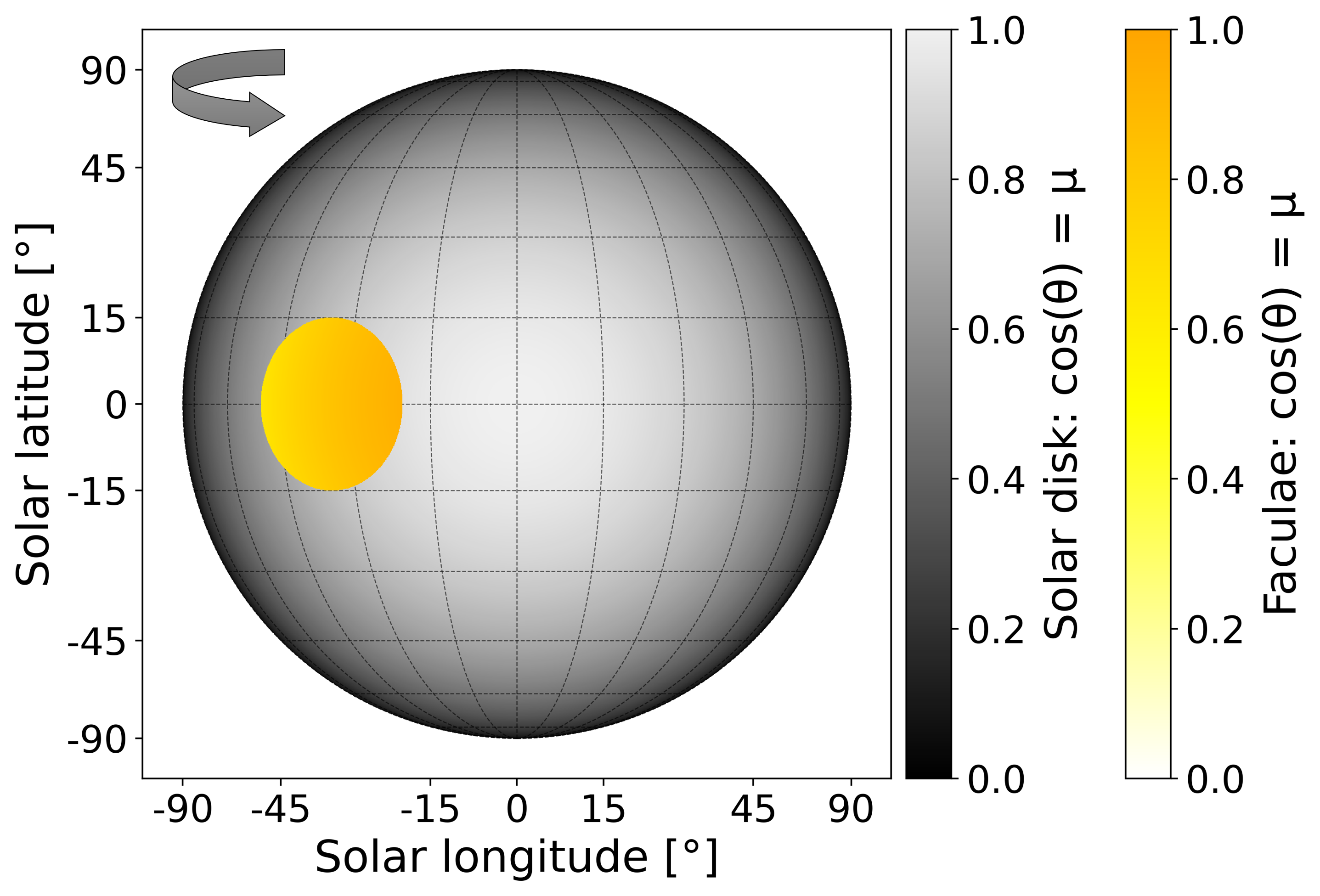}
    \caption{A snapshot from our equatorial facular transit setup. It shows a projected solar disk with a circular patch of faculae (having a radius of 15$^\circ$, represented in yellow shades) located at $-35^\circ$ longitude. The colour scales indicate $\mu$ values.}
    \label{fig:sun_projection}
\end{figure}

To model the transit of a circular patch of faculae across the solar equator (as shown in Figure~\ref{fig:sun_projection}), we decompose the solar surface into pixels on the longitude--latitude grid. In this pixel-based approach, at any given time during the transit, each pixel is assigned either a QS or a facular spectrum depending on whether it lies within the transiting facular patch or not. The prescribed circular facular region is therefore represented on the discrete grid as a pixelized mask; details of its numerical implementation are given in Appendix~\ref{subsec:appendix_grid_model}. The disk-integrated flux spectrum is computed by summing over all pixels on the visible hemisphere, accounting for viewing angle changes across the pixels, projection, and geometrical effects (see Appendix~\ref{subsec:appendix_grid_model} for details).

To include solar rotation in this pixel-based framework, we assign each surface pixel a line-of-sight (LOS) rotational velocity according to its projected position on the disk and apply the corresponding Doppler shift to the local spectrum before disk integration, assuming solid-body rotation. We analyse two configurations: a `\emph{rotating case}', in which these rotational Doppler shifts are included when computing the disk-integrated spectra and RVs, and a `\emph{non-rotating case}', in which the facular patch still follows the same prescribed transit across the projected solar disk, but the local spectra are not assigned any rotation-induced LOS velocity shifts. In this way, the geometry and time-dependent visibility of the transit are kept identical between the two cases, and only the spectroscopic effect of rotation is removed. This allows us to isolate and quantify the impact of rotation on the facular transit signal (see Section~\ref{sec:Results}).

The time variability of the disk-integrated flux arises as a result of the change in the position of the facular patch as it transits the solar disk. We use these resulting disk-integrated spectra at each time step, along with a reference QS disk-integrated spectrum (computed in the absence of the facular patch), to isolate the effects of faculae and extract both the flux difference and RV signals induced by the faculae. A detailed exploration of the flux response of the facular patch transit can be found in Appendix~\ref{subsec:appendix_photometry}. The RV analysis is described in Section~\ref{sec:Results}.

\subsection{RV calculations}\label{subsec:RV-methods}

We calculate faculae-induced RV signals in spectral lines using three complementary methods: flux-weighted centre of gravity (COG) shifts, Gaussian-fitted centroid RVs relative to a master spectrum, and a linearised template matching approach \citep{Bouchy2001, Zechmeister2018}. Each method captures line shifts and asymmetries in different ways, yielding slightly different RV amplitudes but consistently recovering the same overall trend. The choice of these three approaches is somewhat arbitrary, as our primary goal is not to fine-tune RV extraction details, but to understand the underlying physical processes that drive RV variability. Averaging the RVs from all three methods therefore provides a balanced estimate that mitigates method-dependent biases, a common and robust practice in RV studies. Further mathematical details of the different RV extraction methods used here are provided in the Appendix~\ref{subsec:appendix_rv_methods}, together with a comparison of the results from the three methods (Figure~\ref{fig:rv_methods_FeI439}).

All RV signals reported in this study are computed relative to the QS reference spectrum, i.e.,\ they represent the faculae-induced RV perturbation relative to an all-QS disk, rather than the absolute disk-integrated (`Sun-as-a-star') RV of the Sun. For simplicity, we label these relative curves as `RV' throughout the figures and text; we reserve the notation `$\Delta$RV' for differences between two model scenarios (e.g.\ rotating minus non-rotating). Positive RV values correspond to redshifts (receding motions), whereas negative RV values correspond to blueshifts (approaching motions).

\section{Results}\label{sec:Results}

The RV signal induced by the transit of the facular patch arises from the interplay of several effects and has a rather sophisticated profile which we showcase in Section~\ref{subsec:full_picture}. We then disentangle the individual processes that shape the RV profile:  solar rotation in Section~\ref{subsec:fac_rv} and convective blueshift (CB) inhibition by facular magnetic fields in Section~\ref{subsec:inhibition}. We discuss the physical origin of the  CB inhibition in Section~\ref{subsec:discussion}. Finally, in Section~\ref{subsec:multi-line}, we quantify how the faculae-induced RV signal varies across spectral lines.

\subsection{The RV signal from the facular transit}\label{subsec:full_picture}

\begin{figure}[ht!]
    \centering
    \includegraphics[width=0.8
        \linewidth]{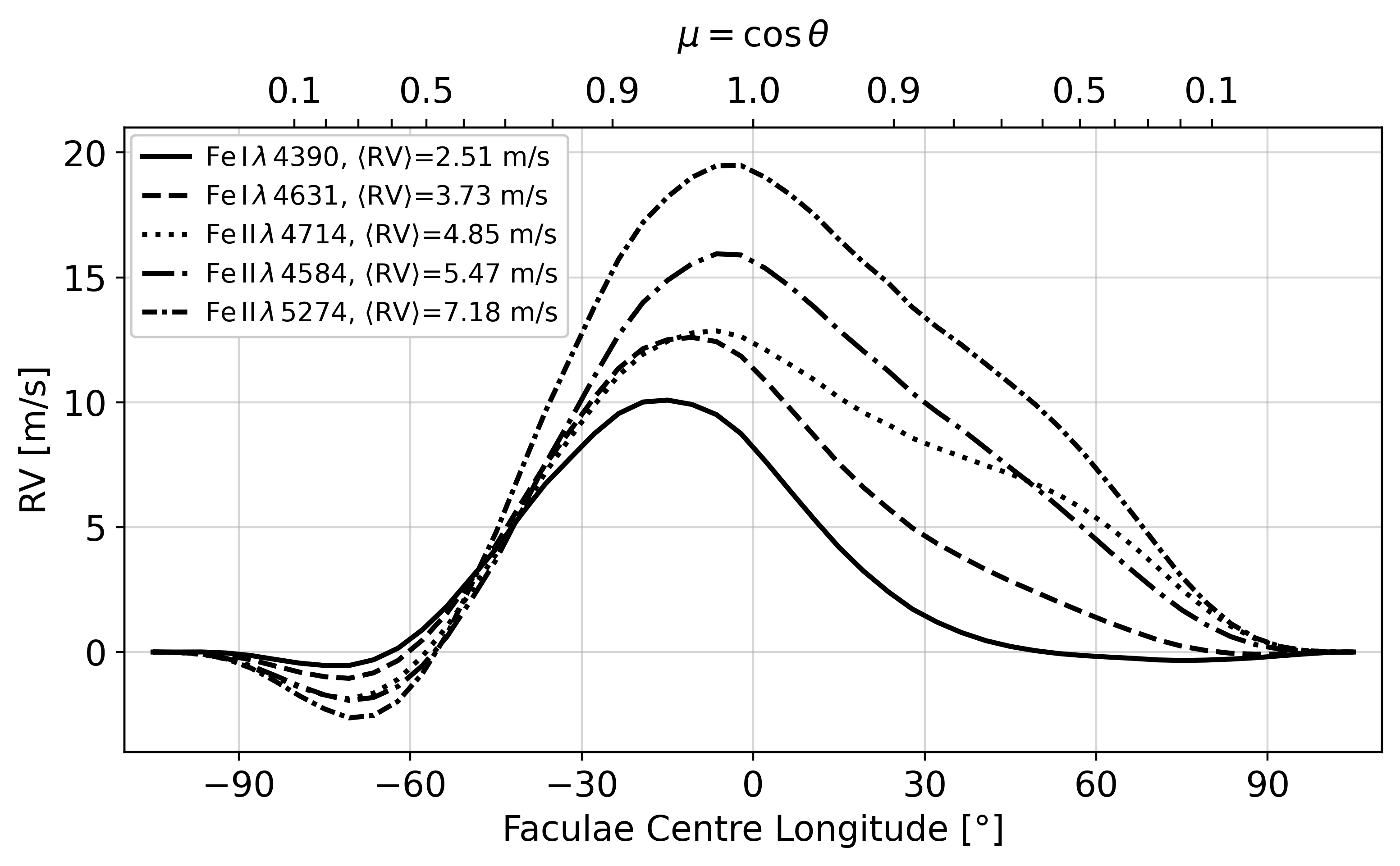}
    \caption{Disk-integrated RV signals induced by a transiting facular patch, as seen in five representative Fe\,I and Fe\,II spectral lines (see Tables \ref{FeI_lines} and \ref{FeII_lines} for line parameters). The RV signals from the `rotating case' are plotted as a function of the longitude of the centre of the facular patch. The longitude range extends to $\pm 105^\circ$ (i.e., $\pm(90^\circ+15^\circ)$) so the $15^\circ$-radius circular patch is initially just beyond the limb and then rotates fully onto and off the visible disk.
    A secondary x-axis at the top indicates the corresponding $\mu$ values. The mean RV over the full transit for each line is given in the legend. We see that all lines exhibit asymmetric RV profiles. The differences between lines increase on the receding hemisphere, while the strongest RV signals occur when the facular patch is still on the approaching hemisphere.}
    \label{fig:full_pic}
\end{figure}

The RV profiles associated with the transit of the facular patch across the solar disk exhibit a complex morphology (Figure~\ref{fig:full_pic}). The profiles are clearly asymmetric, with the maximum RV signal occurring before the facular patch reaches its maximum projected area at central meridian passage. Interestingly, such a 
phase lag (or to be more precise `phase advance') has been inferred in Sun-as-a-star observations with HARPS-N \citep[see Figure 15 from][]{Cameron2019}. During most of the transit, the presence of the facular patch induces a relative redshift. Blueshifts occur shortly after the patch first appears on the eastern limb and in some of the lines also on the western limb before the patch rotates off the visible disk.

Notably, the RV response to faculae is line-dependent: both shapes and amplitudes of the RV profiles strongly change from line to line. This differential behaviour is precisely what distinguishes stellar jitter from Doppler shift caused by reflex motion and, thus, enables the development of the LBL mitigation techniques. Interestingly, the differences in behaviour of the various lines are most noticeable when the facular patch is on the receding hemisphere.

\subsection{The contribution from solar rotation}\label{subsec:fac_rv}

\begin{figure}[t]
    \centering
    \includegraphics[width=0.7
        \linewidth]{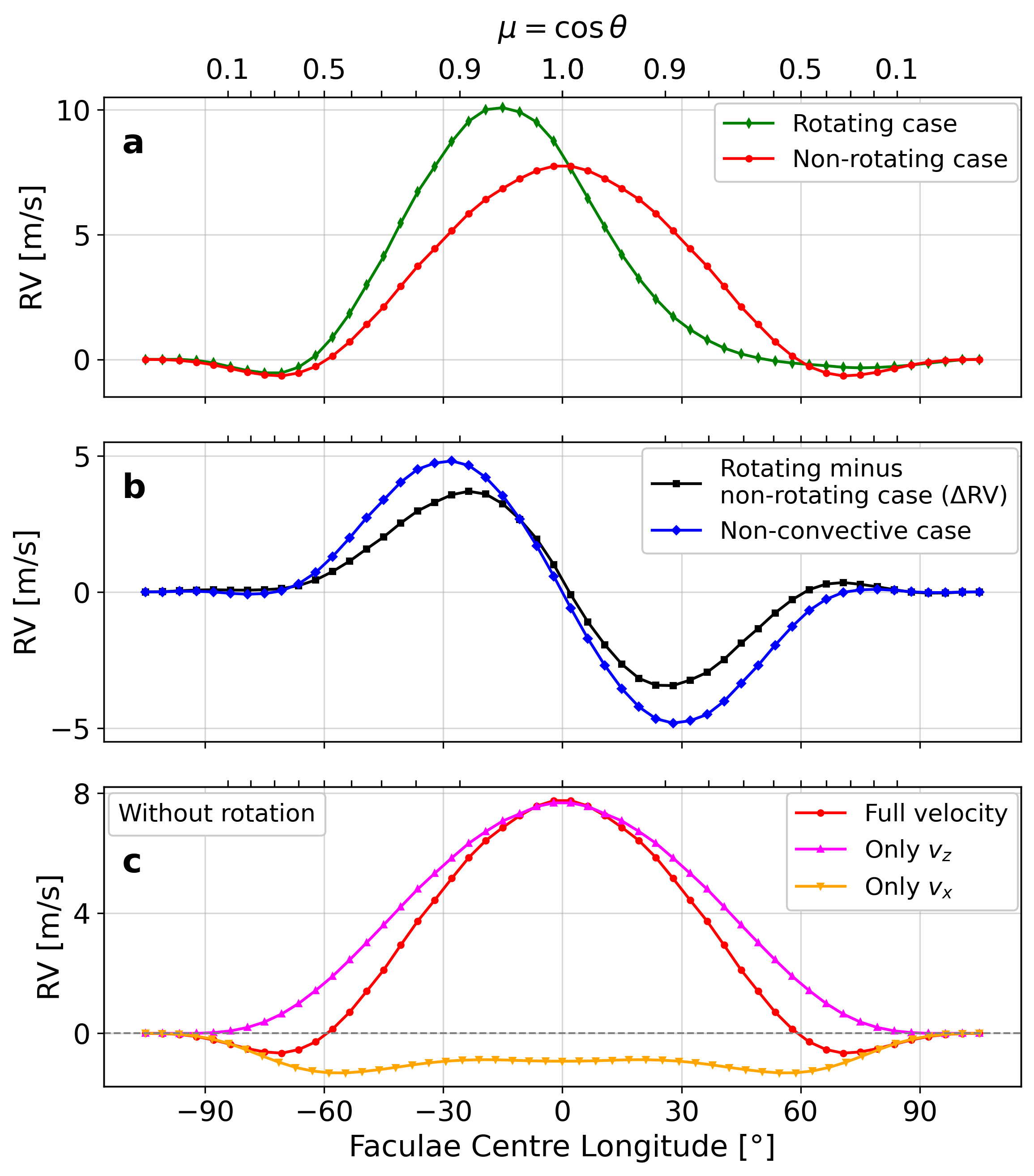}
    \caption{Decomposition of the facular transit RV profile for the Fe\,I\,$\lambda$4390 line. Panel (a): RV profile in the `rotating case' (green, marker: diamonds) is compared to the profile in the `non-rotating case' (red, marker: circles). The green curve is the same as the black solid line in Figure~\ref{fig:full_pic}. Panel (b): difference between the `rotating' and `non-rotating' RV profiles presented in panel (a) (black, marker: squares), plotted along with the RV profile from the `non-convective case' (blue, marker: diamonds), in which the facular thermodynamic and magnetic structure is retained but all convective velocities are set to zero. Panel (c): RV profiles resulting from LOS Doppler shifts due to only vertical convective velocities (only $v_z$; magenta, marker: triangles), due to only horizontal velocities (only $v_x$; orange, marker: upside-down triangles), and due to both horizontal and vertical convective velocities (Full velocity; red, marker: circles). Red curves in panels (a) and (c) are equivalent. See Section~\ref{sec:Results} for further details.}
    \label{fig:M-shape}
\end{figure}

We use the representative Fe\,I\,$\lambda$4390 line (solid curve in Figure~\ref{fig:full_pic}; Table~\ref{FeI_lines} in Appendix~\ref{subsec:appendix_spectral_input}) to explain the physical origin of the complex facular RV morphology in Figure~\ref{fig:full_pic}.

\begin{figure}[ht!]
    \centering
    \includegraphics[width=0.8
        \linewidth]{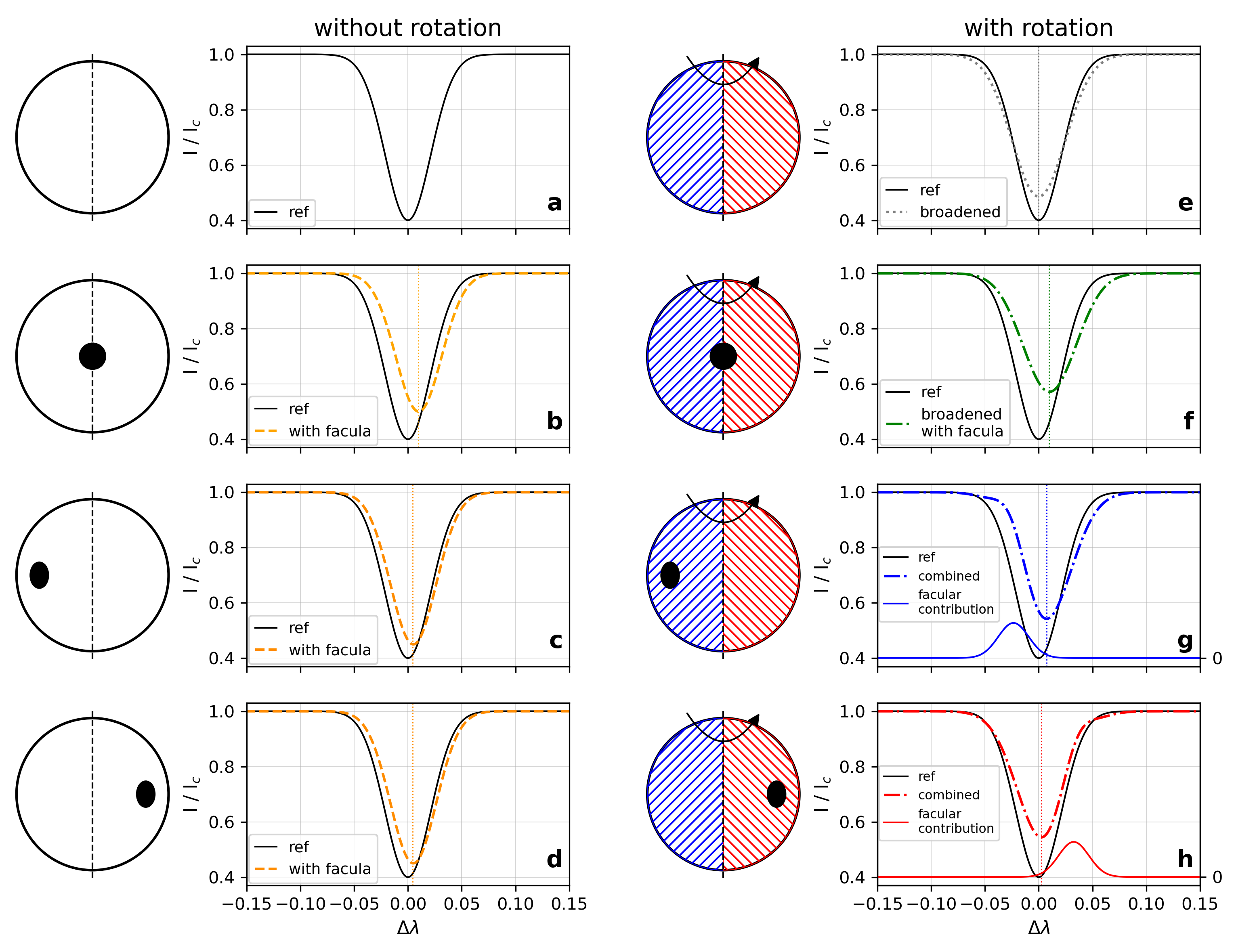}    
        \caption{Schematic illustration of how faculae and solar rotation modify a spectral line profile (line changes are exaggerated for clarity). Panels (a)--(d) show line profiles in the absence of solar rotation, while panels (e)--(h) include rotation. From top to bottom, the rows correspond to: QS only, a facular patch at disk centre, a facular patch on the approaching hemisphere, and a facular patch on the receding hemisphere. The asymmetric QS profile in panel (a) serves as the reference for panels (b)--(h). In panels (g) and (h), the hemisphere-dependent facular contribution is also shown separately (solid, coloured curve) to illustrate its effect on the combined disk-integrated profile. Note: dashed orange curves in panels (c) and (d) are equivalent, whereas dashed-dotted lines in panels (g) and (h) are not (explained in the main text).}
    \label{fig:sketch_line}
\end{figure}

We start with a comparison of the rotating case, presented in Section~\ref{subsec:full_picture}, and the non-rotating case, in which the same facular transit across the disk is retained but the rotation-induced Doppler shifts are set to zero (Section~\ref{subsec:sim_setup}). The comparison reveals that the asymmetry of the RV profiles and the phase lag are attributed to stellar rotation (see Figure~\ref{fig:M-shape}a). Indeed, without rotational Doppler shifts the facular transit imprints an RV signal that is symmetric about central meridian (red curve in Figure Figure~\ref{fig:M-shape}a). This symmetry reflects the axisymmetric nature of the underlying driver: the modification of the convective velocities by magnetic field does not depend on whether the facular patch is on the eastern or the western hemisphere (see Section~\ref{subsec:inhibition} for a detailed discussion). This is showcased by the dashed orange curves in Figure~\ref{fig:sketch_line}c and Figure~\ref{fig:sketch_line}d being identical.

Rotation introduces an inequality between the approaching and receding hemispheres, creating the potential for axisymmetry to be broken. In the absence of magnetic features, rotational blue- and redshifts from the approaching and receding hemispheres cancel in the disk-integrated RV, producing only line broadening (Figure~\ref{fig:sketch_line}e). In the presence of a facular patch, however, that cancellation is no longer exact. Faculae are bright relative to the QS in the cores of most of the spectral lines. This implies that the presence of faculae on the approaching hemisphere brings extra photons to the blue wing, creating an imbalance between the two wings and thus skewing the line to the red (Figure~\ref{fig:sketch_line}g). The opposite behaviour occurs when the patch is on the receding hemisphere, where extra photons are added to the red wing, producing the opposite imbalance and a blueshift of the line (Figure~\ref{fig:sketch_line}h).

Consequently, the rotating case is predominantly redshifted relative to the non-rotating case during the first half of the facular transit and blueshifted during the second half (Figure~\ref{fig:M-shape}a, see also Figure~\ref{fig:trailed_spectra} for the corresponding evolution of the line profile).

To further illustrate the RV signature of rotation, we compute RV profiles while neglecting convective velocities (hereafter, `\emph{non-convective case}'). This is achieved by synthesising spectra from MURaM snapshots with all convective velocities artificially removed, followed by repeating the procedure described in Sections~\ref{subsec:sim_setup}--\ref{subsec:RV-methods}. The resulting RV signal becomes antisymmetric about the central meridian (blue curve in Figure~\ref{fig:M-shape}b): it is positive during the first half of the transit and negative during the second half.

Interestingly, the difference between rotating and non-rotating cases is neither antisymmetric nor identical to the non-convective case (compare blue and black curves in Figure~\ref{fig:M-shape}b). This mismatch highlights that the rotating case is not a simple linear superposition of the symmetric non-rotating case and the antisymmetric non-convective case. The non-linearity arises because magnetic fields modify the local line shape, meaning the spectroscopic effect of rotation is highly sensitive to the presence of underlying convective velocities.

\subsection{The contribution from CB inhibition}\label{subsec:inhibition}

\begin{figure}[t]
    \centering
    \includegraphics[width=0.7
        \linewidth]{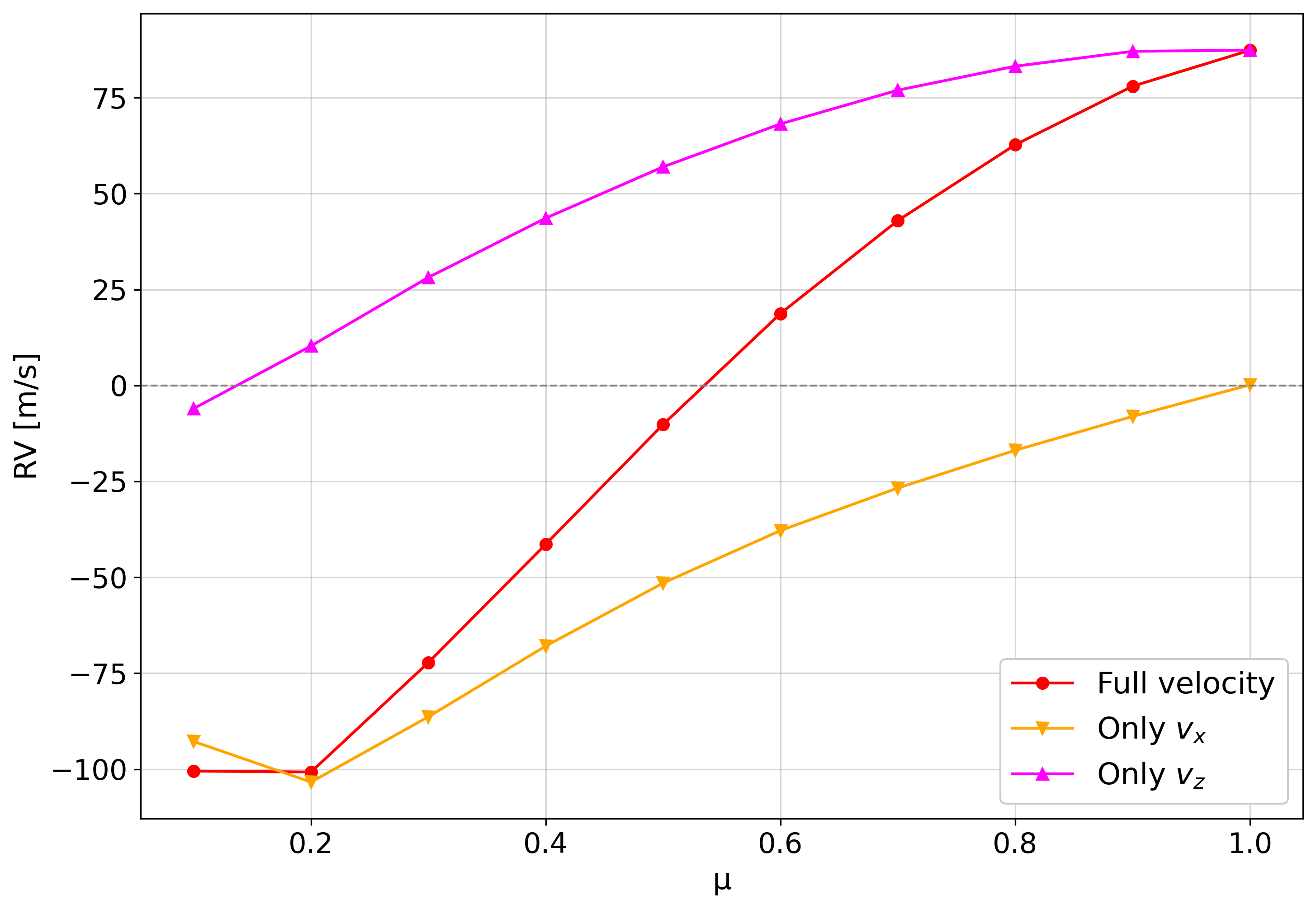}
     \caption{RV signal induced by facular suppression of CB as a function of $\mu$ for the Fe\,I\,$\lambda$4390 line. RV due to LOS Doppler shifts arising only from vertical velocities (magenta), only from horizontal velocities (orange), and both vertical and horizontal convective velocities (red). All RVs are computed from raw $\mu$-dependent facular spectra, using the corresponding QS spectrum as the reference.}
    \label{fig:µ-dep_inhibition}
\end{figure}
In the next step, we investigate what physical phenomena contribute to the shape of the RV profile of the non-rotating case (red curve in Figure~\ref{fig:M-shape}a). In contrast to the effect of stellar rotation, which is caused by the motion of the star as a whole, this effect is due to convective motions in the stellar atmosphere. They distort the spectral lines, and thereby affect the measured velocity of a star. The presence of magnetic fields in facular regions modifies these convective flows, leading to an apparent difference between the velocities measured in QS and facular atmospheres.

While the non-rotating case results in a symmetric RV profile with its maximum signal (highest redshift) corresponding to the facular patch crossing the central meridian, the overall shape of the profile is rather complex. In particular, while the facular patch mostly leads to a positive RV (redshift due to the inhibition of CB by magnetic field) over the course of the transit, the RV surprisingly appears negative (blueshift) at the beginning and at the end of the transit when the facular patch is close to the limb. 

To explore the origin of this behaviour, we performed two experiments. In the first experiment, we repeated the spectral synthesis (see Section~\ref{subsec:spectral_input}) ignoring the horizontal component of convective velocities and only accounting for Doppler shifts caused by vertical velocities (hereafter, referred as `\emph{only} $v_z$' case). In the second experiment, we ignored vertical velocities and retained only Doppler shifts from the horizontal component (hereafter, referred as `\emph{only} $v_x$' case). We used the same Fe\,I\,$\lambda$4390 line as in Figures~\ref{fig:mu_spectra} and \ref{fig:M-shape} and followed the methodology explained in Section~\ref{subsec:RV-methods} to measure the RV signal between faculae and QS spectra as a function of disk position. The results are shown in Figure~\ref{fig:µ-dep_inhibition}. The `only $v_z$' case leads to a redshift except very close to the limb where reversed granulation \citep[see, e.g.][]{Cheungetal2007} leads to a small blueshift. The `only $v_x$' case yields a blueshift at all disk positions, except at disk centre where horizontal velocities have no LOS component. In other words, instead of CB inhibition, horizontal velocities lead to CB enhancement.

The RV signal (red curve in Figure~\ref{fig:µ-dep_inhibition}) is a superposition of the `only $v_x$' and `only $v_z$' signals. At high $\mu$-values, the signal is `only $v_z$'-dominant and, thus, positive. In contrast, low $\mu$-values correspond to an `only $v_x$'-dominant case and a negative signal. The transition between positive and negative RV values occurs around $\mu \approx 0.5$ for the Fe\,I\,$\lambda$4390 line shown in Figure~\ref{fig:µ-dep_inhibition}, although its exact location is strongly line-dependent (see Figure~\ref{fig:µ-dep_inhibition_multi-line}). 

The dependences plotted in Figure~\ref{fig:µ-dep_inhibition} result in the corresponding transit profiles shown in Figure~\ref{fig:M-shape}. We note that the amplitudes of the RV signals shown in these figures are very different. While the value of the relative RV signal between faculae and QS spectra ranges from approximately -100 m/s to 90 m/s (Figure~\ref{fig:µ-dep_inhibition}), the facular-transit experiment returns substantially smaller values between -0.5 m/s and 8 m/s (Figure~\ref{fig:M-shape}), because it is diluted by disk integration. The dilution is much higher closer to the limb than at the disk centre due to foreshortening and limb darkening. Consequently, the signal from horizontal velocities (dominating near-limb regions) is more strongly affected than that from vertical velocities (dominating central-disk regions).

Horizontal velocities have no LOS component at disk centre, so both the QS and facular spectra at disk centre are unshifted with respect to the laboratory frame in the 'only $v_x$ case'. Nevertheless, we observe a non-zero RV signal when the facular region crosses the central meridian (Figure~\ref{fig:M-shape}c). This arises because the RV is measured relative to the disk-integrated QS spectrum, which is blueshifted with respect to the laboratory frame even in the 'only $v_x$ case' (see above).

\subsection{The origin of CB inhibition}\label{subsec:discussion}

To better understand the behaviour in Figure~\ref{fig:µ-dep_inhibition}, we first examine the physical origin of CB in the quiet Sun and, in particular, the respective roles of vertical and horizontal convective velocities. We then discuss how these processes are modified in facular regions.

\begin{figure}[ht]
    \centering
    \includegraphics[width=0.6
        \linewidth]{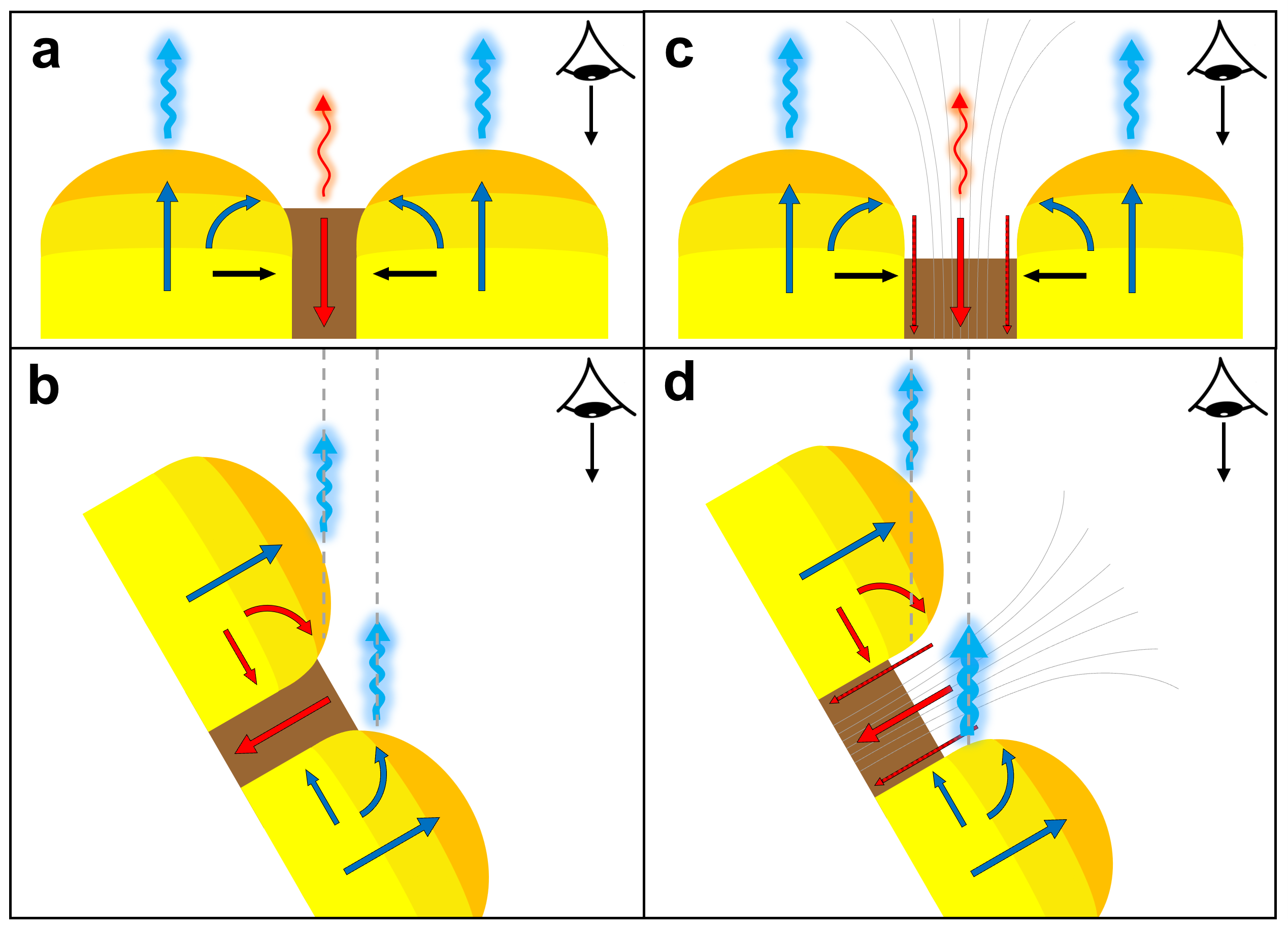}
    \caption{Cartoon illustrating how surface corrugation affects the flux-weighted RV as a function of $\mu$ in quiet Sun and faculae. Panels (a) and (b) show the QS case at disk centre ($\mu=1$) and for an inclined view ($\mu=0.5$), respectively. Panels (c) and (d) show the corresponding facular cases with strong magnetic concentrations in intergranular lanes (grey field lines) and a Wilson depression. Granules are shaded orange--yellow (hotter towards yellow) and intergranular lanes are brown. Straight arrows denote convective flows and wavy arrows emergent intensity; blue (red) indicates LOS velocity components producing blue- (red-) shifted contributions, while black arrows have negligible LOS projection. In panels (c) and (d), the hatched red arrows mark the narrow high-speed downflow lanes adjacent to the magnetic concentration. The eye symbol and arrow in each panel indicate the viewing direction (LOS). The dashed lines are included as a visual guide to the LOS geometry: owing to the 3D corrugated $\tau\!\approx\!1$ surface, parts of the near-side granular wall are occulted at inclined viewing angles, while deeper layers on the far-side wall can become visible. In faculae, magnetic pressure enhances this effect via the Wilson depression (`hot-wall' visibility), thereby changing the flux weighting of upflows and downflows. Consequently, the convective blueshift is not uniformly suppressed with decreasing $\mu$; depending on disk position and magnetic structuring, the net Doppler signal can be inhibited or, close to the limb, even enhanced (see Section~\ref{subsec:inhibition}).}
    \label{fig:facula_sketch}
\end{figure}

The CB due to vertical velocities represents the textbook case. It arises from an intrinsic asymmetry of convective motions imposed by the need to transport energy outward against gravity. This asymmetry manifests itself in convective upflows being generally hotter (and therefore brighter) than downflows (see Figure~\ref{fig:facula_sketch}a). The resulting positive correlation between temperature and velocity leads to the CB \citep[pp.~418--424]{Gray2022}.

The effect of horizontal velocities is more intricate, since convective motions in the horizontal directions are symmetric: all horizontal directions are statistically equivalent, so that there is no net horizontal velocity associated with convective motion. Horizontal velocities still contribute to the blueshift (away from disk centre) due to the corrugation of the optical surface. It interferes with the visibility of flows towards and away from the observer, breaking the symmetry between them. Specifically, granules are elevated relative to intergranular lanes, so that the near-side of granules with positive projections of horizontal velocities to the LOS are better visible than far-sides with negative projections \citep[see also the description in][]{Yuetal2026}. We show this in Figure~\ref{fig:facula_sketch}b by considering two symmetric points with opposite projections of horizontal velocity to the LOS (grey dashed lines). The point with a positive component (corresponding to blueshift) is exposed to the observer while the point with a negative component (corresponding to redshift) is obscured by the top of the granule (which, depending on the exact geometry, can even induce a blueshift as in the case shown in Figure~\ref{fig:facula_sketch}b).

Having discussed the operation of near-surface convection and the origin of CB in the quiet Sun, we now examine how these processes are modified in facular regions. Faculae are regions on the solar surface occupied by concentrations of small-scale, predominantly vertical kG magnetic fields located in intergranular lanes (see Appendix~\ref{subsec:appendix_spectral_input}). First, the vertical magnetic field inhibits horizontal convective flows reducing efficiency of the convection.
Second, it affects vertical flows but in a rather counter-intuitive way by increasing their velocity. In particular, it increases velocities in downflows in the vicinity of kG magnetic concentrations \citep{Beeck2015}, and observations with high spatial resolution even revealed strong, sometimes supersonic downflows, which in mature plages are often found in narrow lanes surrounding the magnetic concentrations \citep[][]{Shimizuetal2008, Buehleretal2015}.

Third, it alters the surface morphology. Magnetic flux concentrations \citep[often referred in the literature as flux tubes, e.g.,][]{Solanki1993} can widen intergranular lanes and squeeze adjacent granules. They also enhance surface corrugation due to the effect known as the Wilson depression \citep[][see Figure~\ref{fig:facula_sketch}c]{Solankietal2006}. This depression is caused by an increase of the magnetic pressure in the magnetic flux concentrations, which in turn leads to a decreased gas pressure and density. This evacuation leads to an opacity decrease in the intergranular magnetic flux concentrations. In other words, the visible surface is geometrically depressed inside the magnetic flux concentrations. This depression exposes the deeper and hotter sidewalls of the surrounding granules (see Figure~\ref{fig:facula_sketch}c), setting up the hot-wall effect which becomes prominent towards the limb as the LOS passes obliquely through the depression and pierces the hot walls (see Figure~\ref{fig:facula_sketch}d). At disk centre one looks straight down into the depression, which might be dark or bright depending on the wavelength (Figure~\ref{fig:mu_spectra}, Figure~\ref{fig:trailed_spectra}). 

Larger integranular lanes with more vigorous (relative to the non-magnetic case) downflows shift spectral lines to the red, i.e., it causes positive RV of faculae relative to the quiet Sun. The only exception is the limb region ($\mu=0.1$) where the RV signal arising exclusively from vertical flows becomes negative (magenta curve in Figure~\ref{fig:µ-dep_inhibition}) because the granulation pattern reverses in the higher photospheric layers from which the limb radiation emerges.

The effect of magnetic field on the CB caused by horizontal flows is more intricate. Despite the overall decrease in horizontal velocities, the blueshift associated with them increases. This is because the Wilson depression enhances the visibility of granular parts that move towards the observer (by exposing deeper layers of granules, see Figure~\ref{fig:facula_sketch}d), thus, increasing their projected area. These parts are also hotter (and brighter) and have larger horizontal velocities than the parts visible in the non-magnetic case. Thus, the resulting RV signal is negative and its absolute value increases towards the limb. We note that the monotony is lost at the limb ($\mu=0.1$) which might be due to geometric occultation: at very shallow viewing angles, foreground granulation partially hides the hot-wall region \citep[see, e.g.,][]{Albertetal2025}.

The discussion presented in this section is rather schematic. The exact distribution of convective velocities in the solar photosphere is very complicated (see Figures~\ref{fig:muram_6} and \ref{fig:muram_3} in Appendix~\ref{subsec:appendix_flow_physics}). It might also depend on the evolutionary phase of the flux tube, i.e.\ whether the field is still being amplified by the convective collapse or the flux tube is already decaying. The detailed quantitive analysis of the physical mechanisms defining the effect of the magnetic field on the CB will be given in a forthcoming paper.

\subsection{Differential RV}\label{subsec:multi-line}

\begin{figure}[t]
    \centering
    \includegraphics[width=0.6
        \linewidth]{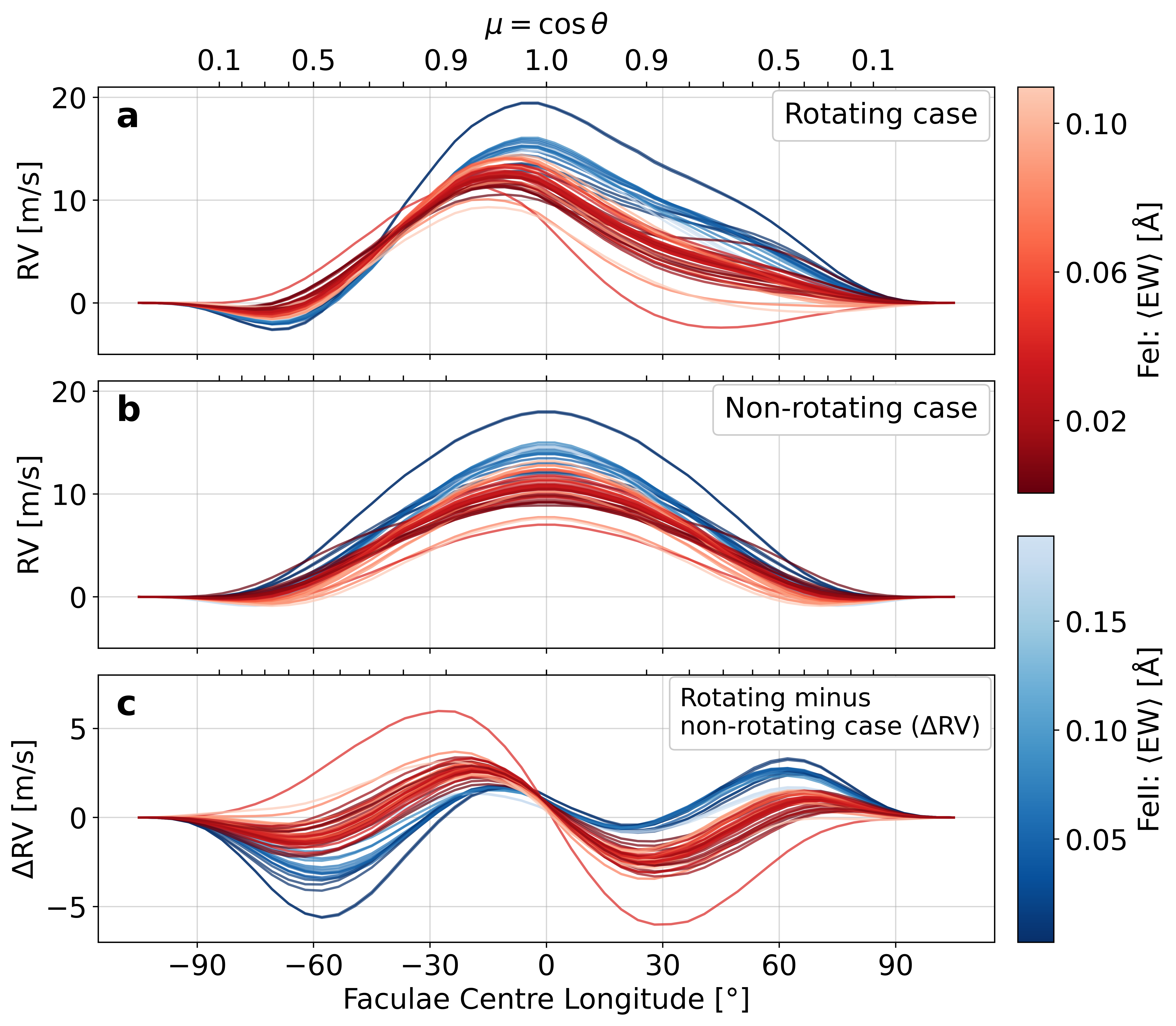}
    \caption{Facular transit RV profiles for 60 perfectly isolated (blend-free) Fe\,I (red shades) and Fe\,II (blue shades) lines. Panel (a): rotating case, panel (b): non-rotating case, and panel (c): difference between the rotating and non-rotating cases, highlighting the line-dependent rotation--faculae coupling (so that it is not a pure `rotation-only' signal; see Section~\ref{subsec:fac_rv}). The RV profiles are colour coded according to the line strength (mean equivalent width (EW) over the transit) as indicated by the colour bars.}
    \label{fig:M-shape_multi-line}
\end{figure}

We now address the faculae-induced RV signals across all lines from Fe\,I and Fe\,II species in our sample (see Figure~\ref{fig:M-shape_multi-line}, Tables~\ref{FeI_lines} and~\ref{FeII_lines}, and also Appendix~\ref{subsec:appendix_mu_inhibition}). The spectral lines clearly exhibit a highly differential response to the presence of faculae. The shape, the amplitude, and the asymmetry of the RV signal (including phase lag) differ substantially between lines within and across the two species (see Figure~\ref{fig:M-shape_multi-line}a). Interestingly, the differences between RV signals measured in different lines appear to be more pronounced on the receding hemisphere, as already indicated by Figure~\ref{fig:full_pic}.

The differences between the RV responses of spectral lines arise mainly from the combination of three effects. First, the contribution from solar rotation (Section~\ref{subsec:fac_rv}) depends primarily on the facular contrast in a given line. Second, the influence of the magnetic field on the correlation between temperature and vertical velocity, which governs the inhibition of convective blueshift, is strongly depth-dependent. Third, the contribution from horizontal velocities depends on the corrugation of the optical surface in the line and the nearby continuum.

These three effects depend on different line properties. In particular, the latter two are sensitive to the atmospheric layers sampled by the line, whereas the rotational contribution is controlled mainly by the line-dependent facular contrast. The line response is further shaped by the temperature sensitivity of the line opacity, which is governed by the Saha-Boltzmann distribution and therefore depends on both the ionisation balance and the excitation energy of the lower level above the ground state. This naturally leads to different temperature responses in Fe\,I and Fe\,II lines. In addition, the magnitude of the response depends on line strength, with saturated lines being less temperature-sensitive than weaker lines.

Such differential sensitivity of spectral lines to faculae stands in stark contrast to the spectral signature of stellar reflex motion, where all lines respond identically. It thus paves the way for disentangling planetary and stellar faculae-driven signals.

\section{Conclusions}\label{sec:summary}

Using disk-resolved quiet-Sun and facular spectra synthesised with the MPS-ATLAS code from 3D radiative MHD MURaM simulations, we modelled the RV signature of a single facular patch transiting the visible disk. To our knowledge, this is the first ever study of faculae-induced RV signal based on 3D radiative MHD simulations.

At disk centre, faculae produce a relative redshift because magnetic fields modify near-surface convection (the classic convective blueshift suppression). 
Toward the limb, the facular effect switches to a relative blueshift for many lines, with the transition typically occurring near $\mu \approx 0.5$ (its exact location is line-dependent). This blueshift is caused by the enhancement of the corrugation of the optical surface by the magnetic field, which results in better visibility of horizontal flows along the LOS.

In the absence of rotation-induced Doppler shifts, the facular-transit RV profile is symmetric about its central meridian crossing. Including rotational Doppler shifts produces an asymmetric RV curve and a phase lag, with the maximum RV occurring before the maximum projected area, consistent with the qualitative behaviour inferred from Sun-as-a-star observations \citep[e.g., ][]{Cameron2019}.

Across 60 isolated Fe\,I and Fe\,II lines considered in this study, both the amplitude and morphology of the RV signal, including its asymmetry and phase lag, vary substantially from line to line. This differential response contrasts with stellar reflex motion, which shifts all lines identically, and provides a physically grounded basis for line-by-line activity mitigation and optimised line weighting.

\nolinenumbers

\begingroup
\let\internallinenumbers\relax
\let\endinternallinenumbers\relax

\begin{acknowledgements}

F.K., K.S., A.I.S., A.C.C., and V.W.\ acknowledge support from the European Research Council (ERC) under the European Union’s Horizon 2020 Research and Innovation Programme through grant No.\ 101118581. S.K.S.\ acknowledges support from the ERC under the same programme through grant No.\ 101097844.
I.R.\ acknowledges financial support from the European Research Council (ERC) under the European Union’s Horizon Europe programme (ERC Advanced Grant SPOTLESS; No.\ 101140786), from Spanish grants PID2021-125627OB-C31 and PID2024-158486OB-C31 funded by MCIU/AEI/10.13039/501100011033 and by “ERDF A way of making Europe”, from the programme Unidad de Excelencia María de Maeztu CEX2020-001058-M, from the Generalitat de Catalunya/CERCA programme, and from the Catalan Government via the SGR 01526/2021 grant.
We gratefully acknowledge the computational resources provided by the Raven supercomputer systems of the Max Planck Computing and Data Facility (MPCDF) in Garching, Germany.
This study has made use of SAO/NASA Astrophysics Data System’s bibliographic services.

\end{acknowledgements}

\bibliography{papers}

\appendix
\restartappendixnumbering

\makeatletter
\@addtoreset{figure}{section}
\@addtoreset{table}{section}

\@ifundefined{theHfigure}{}{%
  \renewcommand{\theHfigure}{appendix.\thesection.\arabic{figure}}%
}
\@ifundefined{theHtable}{}{%
  \renewcommand{\theHtable}{appendix.\thesection.\arabic{table}}%
}
\makeatother

\section{MURaM and MPS-ATLAS Spectral Input}\label{subsec:appendix_spectral_input}

We use two sets of MURaM 3D MHD simulations of the solar photosphere, each covering an area of 9000 x 9000\,km$^2$ (with 512 x 512 spatial pixels) on the solar surface and capturing the evolution of the solar photosphere over a duration of $\sim1$ hour with 90\,s cadence. The first set consists of 45 snapshots from the QS simulations including a small-scale dynamo (SSD) driven by near-surface convection \citep{VoeglerandSchuessler2007,SchuesslerandVoegler2008,Rempel2014}, which are described in detail in \citet{Witzke2024}. The second set, which represents faculae, consists of 40 snapshots of the simulation in which a mean vertical field of 200\,G is introduced to the SSD setup as described in \citet{Witzke2022,Kostogryz2024}, which leads to the formation of small-scale kG flux concentrations in the intergranular lanes.

We note that pressure-modes are self-consistently generated in the simulations \citep[see e.g.][]{Beecketal2012,Jessetal2012,Ceglaetal2013,Frame2025}, which have a typical oscillation period of 5-6 minutes. These modes cause vertical oscillations of the simulated plasma which introduce alternate red- and blueshifts in the emergent spectra. Furthermore, the overall centre of mass of the simulated domain is seen to have a horizontal drift that is a consequence of the horizontal periodic boundary conditions and model initialisation \citep{Frame2025}. Mean horizontal velocities in QS simulations, when averaged over the short time interval of $\sim1$ hour, show a net blueshift of order -100\,m/s, while the faculae simulations show a net redshift of order 200\,m/s. These shifts therefore add an apparent RV contribution that arises from the limited temporal coverage of the simulations as well as due to having a limited number of granules in the simulated domain as compared to about a million granules on the solar disk. To remove the effects of horizontal drifts and vertical oscillations, we perform spectral calculations taking the following approach.

In step 1, we rotate every snapshot of the QS and faculae simulations about a pivot axis at the mean optical surface and generate atmospheres for ten viewing angles ($\mu = 0.1$ -- $1.0$ in steps of $0.1$). Then, we compute spectra emerging from these atmospheres at all ten viewing angles using the MPS-ATLAS radiative transfer code \citep{Witzke2021}, taking into account the convective velocities in the simulations and the resulting Doppler shifts. For the spectral synthesis, we consider a set of 60 fully blend-free Fe\,I and Fe\,II lines (see Tables~\ref{FeI_lines} and~\ref{FeII_lines}) following \citet{Sowmyaetal2025}. We restrict the sample to iron lines because of their abundance and diagnostic utility for Sun-like stars. To accurately capture the subtle changes in line shapes and wavelength shifts, we compute the line profiles at a resolving power of 2,000,000. This high spectral resolution exceeds that of most modern spectrographs, allowing them to be readily adapted to any instrument of choice. The emergent spectra for each viewing angle are then averaged spatially (over all 512 x 512 pixels in the simulation snapshot). This results in 45 QS spectra and 40 facular spectra per $\mu$ and per spectral line. To remove p-modes from the spectra, for each $\mu$ and each spectral line, we carry out a temporal average of all 45 QS and 40 facular spectra.

The resultant spectra from step 1 still contain velocity shifts arising from horizontal drifts in the simulations. Next, in step 2, we rotate the simulation snapshots in a direction opposite to that used in step 1 and prepare the atmospheres for the ten $\mu$ angles (Figure~\ref{fig:v_x_drift_cartoon}). Following the same procedure as in step 1, we compute the emergent spectra at every pixel in every snapshot and then perform spatial and temporal averaging. The resultant spectra will now have velocity shifts from horizontal drifts, which are in opposite direction to those from step 1. We remove the horizontal drift component by taking an average of the spectra from step 1 and step 2 (we note that the line broadening caused by this averaging is negligible). This results in a total of ten spatially and temporally averaged spectra (one at each $\mu$ value) each for QS and faculae for each spectral line, which are free from oscillations and horizontal drifts.

\begin{figure}[ht!]
    \centering
    \includegraphics[width=0.6
        \linewidth]{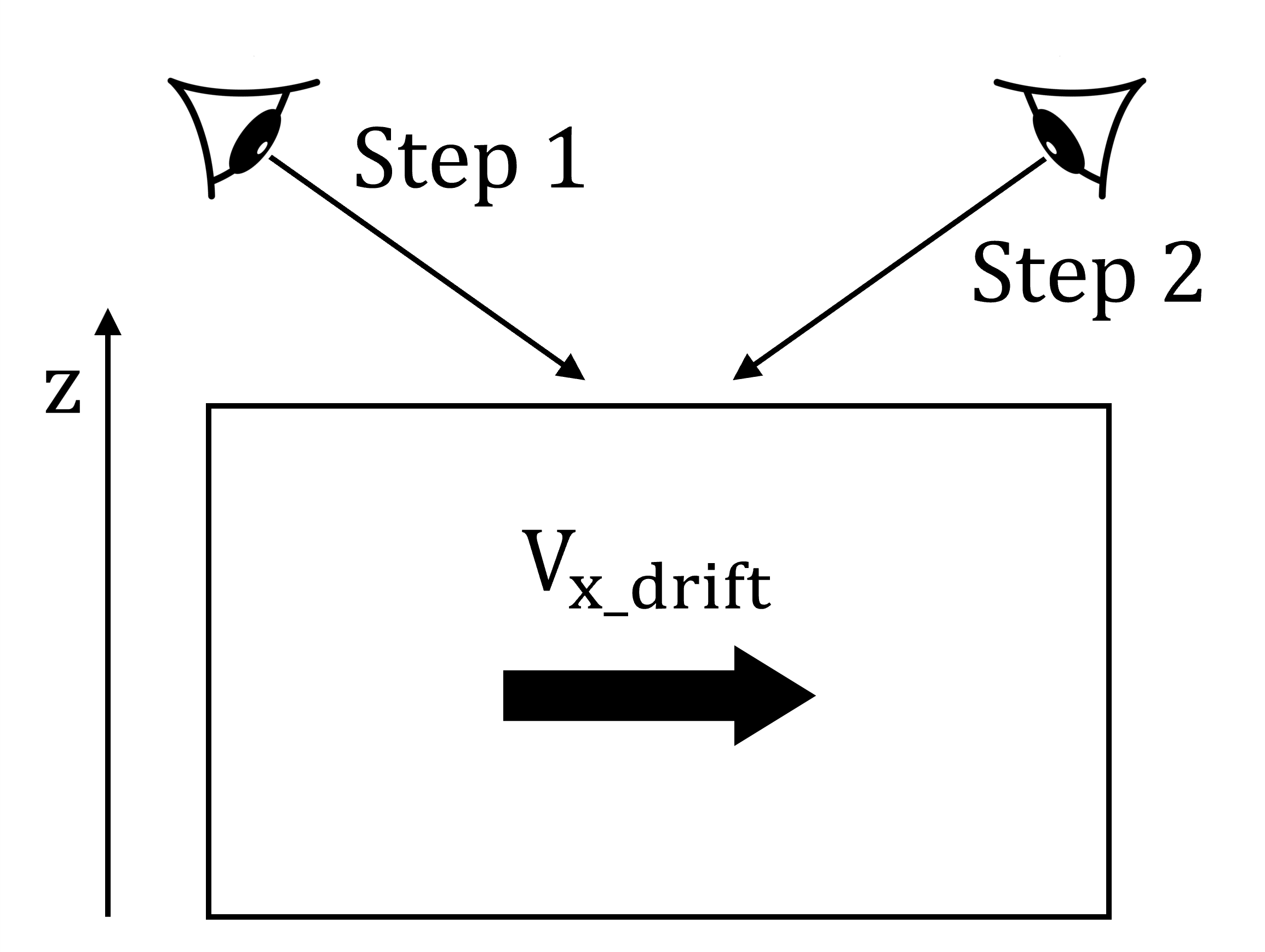}
    \caption{Schematic illustration of the procedure used to remove horizontal drift velocities from the simulated spectra. The rectangle represents a simulation cube, while the horizontal arrow labelled $V_{x\_drift}$ denotes the net horizontal drift present in the snapshots. Emergent spectra are computed by observing the cube from two opposite azimuthal viewing directions (indicated by the upper arrows), corresponding to the original orientation (step 1) and a rotated configuration with reversed horizontal velocity projection (step 2). Averaging the spectra obtained from these two opposite viewing geometries cancels the contribution from horizontal drifts, while preserving the physical line asymmetries and Doppler shifts of interest.}
    \label{fig:v_x_drift_cartoon}
\end{figure}

Finally, we remark that although our spectral synthesis does not include the treatment of the Zeeman effect, this omission is not expected to significantly affect the RV signal induced by faculae and our conclusions from it. Zeeman broadening is only significant when magnetic splitting exceeds the intrinsic line width \citep[see e.g.][]{Beecketal2015}, but this condition is not met for the spectral lines in our selected sample.

\begin{table}[htbp]
    \centering
    \caption{List of the 42 Fe\,I lines used in this study; synthesised with MPS-ATLAS. $E_{\rm low}$ is the excitation energy of the lower level, $\log(gf)$ is the logarithm of the oscillator strength times the multiplicity of the lower level. The atomic parameters are taken from the VALD3 database.}
    \begin{tabular}{cccc}
    \toprule
    \midrule
    Line ID & Vacuum wavelength [\AA] & $E_{\rm low}$ [eV] & $\log(gf)$ \\
    \midrule
    Fe\,I\,$\lambda$4390 & 4390.47666 & 0.052 & -4.583 \\
    Fe\,I\,$\lambda$4420 & 4425.07750 & 3.651 & -1.610 \\
    Fe\,I\,$\lambda$4440 & 4440.87659 & 4.543 & -3.337 \\
    Fe\,I\,$\lambda$4446 & 4446.71574 & 0.087 & -5.441 \\
    Fe\,I\,$\lambda$4480 & 4485.47222 & 3.601 & -0.864 \\
    Fe\,I\,$\lambda$4520 & 4524.66603 & 3.651 & -1.990 \\
    Fe\,I\,$\lambda$4547 & 4547.73708 & 4.182 & -2.510 \\
    Fe\,I\,$\lambda$4557 & 4557.91535 & 3.939 & -2.060 \\
    Fe\,I\,$\lambda$4584 & 4584.21902 & 2.840 & -2.879 \\
    Fe\,I\,$\lambda$4588 & 4588.99826 & 3.975 & -2.150 \\
    Fe\,I\,$\lambda$4594 & 4594.80241 & 3.939 & -2.060 \\
    Fe\,I\,$\lambda$4597 & 4597.69805 & 3.651 & -2.320 \\
    Fe\,I\,$\lambda$4599 & 4599.39951 & 3.276 & -1.570 \\
    Fe\,I\,$\lambda$4603 & 4603.28304 & 1.605 & -3.154 \\
    Fe\,I\,$\lambda$4608 & 4608.93012 & 3.975 & -3.242 \\
    Fe\,I\,$\lambda$4620 & 4620.57774 & 3.601 & -1.120 \\
    Fe\,I\,$\lambda$4628 & 4628.83745 & 3.299 & -3.059 \\
    Fe\,I\,$\lambda$4631 & 4631.41180 & 2.277 & -2.587 \\
    Fe\,I\,$\lambda$4680 & 4680.14801 & 3.601 & -0.833 \\
    Fe\,I\,$\lambda$4727 & 4727.45842 & 2.995 & -3.250 \\
    Fe\,I\,$\lambda$4730 & 4730.33358 & 4.068 & -1.614 \\
    Fe\,I\,$\lambda$4751 & 4751.26891 & 4.554 & -1.340 \\
    Fe\,I\,$\lambda$4795 & 4795.29731 & 3.043 & -3.530 \\
    Fe\,I\,$\lambda$4811 & 4811.28263 & 3.568 & -2.720 \\
    Fe\,I\,$\lambda$4903 & 4903.53164 & 3.229 & -3.866 \\
    Fe\,I\,$\lambda$4906 & 4906.49427 & 3.921 & -2.050 \\
    Fe\,I\,$\lambda$4909 & 4909.09540 & 3.423 & -1.840 \\
    Fe\,I\,$\lambda$4911 & 4911.68808 & 4.182 & -0.461 \\
    Fe\,I\,$\lambda$4926 & 4926.13983 & 2.277 & -2.241 \\
    Fe\,I\,$\lambda$4928 & 4928.79080 & 3.568 & -2.073 \\
    Fe\,I\,$\lambda$4993 & 4993.25335 & 4.211 & -1.910 \\
    Fe\,I\,$\lambda$4995 & 4995.52080 & 0.915 & -3.080 \\
    Fe\,I\,$\lambda$5000 & 5000.49881 & 4.182 & -1.740 \\
    Fe\,I\,$\lambda$5014 & 5014.08858 & 4.279 & -1.790 \\
    Fe\,I\,$\lambda$5017 & 5017.87063 & 4.250 & -1.690 \\
    Fe\,I\,$\lambda$5021 & 5021.12326 & 3.975 & -2.126 \\
    Fe\,I\,$\lambda$5105 & 5105.60854 & 4.172 & -1.970 \\
    Fe\,I\,$\lambda$5111 & 5111.06423 & 4.299 & -0.980 \\
    Fe\,I\,$\lambda$5131 & 5131.05681 & 3.939 & -1.850 \\
    Fe\,I\,$\lambda$5237 & 5237.65366 & 4.182 & -1.497 \\
    Fe\,I\,$\lambda$5264 & 5264.33914 & 3.246 & -2.660 \\
    Fe\,I\,$\lambda$5296 & 5296.02009 & 3.634 & -2.860 \\
    \bottomrule
    \end{tabular}
    \label{FeI_lines}
\end{table}

\begin{table}[htbp]
    \centering
    \caption{List of the 18 Fe\,II lines used in this study; synthesised with MPS-ATLAS. $E_{\rm low}$ is the excitation energy of the lower level, $\log(gf)$ is the logarithm of the oscillator strength times the multiplicity of the lower level. The atomic parameters are taken from the VALD3 database.}
    \begin{tabular}{cccc}
    \toprule
    \midrule
    Line ID & Vacuum Wavelength [\AA] & $E_{\rm low}$ [eV] & $\log(gf)$ \\
    \midrule
    Fe\,II\,$\lambda$4492 & 4492.65471 & 2.853 & -2.700 \\
    Fe\,II\,$\lambda$4584 & 4584.10900 & 2.840 & -3.090 \\
    Fe\,II\,$\lambda$4621 & 4621.80697 & 2.827 & -3.240 \\
    Fe\,II\,$\lambda$4658 & 4658.27748 & 2.886 & -3.610 \\
    Fe\,II\,$\lambda$4668 & 4668.05146 & 2.827 & -3.368 \\
    Fe\,II\,$\lambda$4671 & 4671.46972 & 2.578 & -4.059 \\
    Fe\,II\,$\lambda$4714 & 4714.49464 & 2.776 & -5.247 \\
    Fe\,II\,$\lambda$4721 & 4721.45836 & 3.194 & -4.822 \\
    Fe\,II\,$\lambda$4925 & 4925.29260 & 2.886 & -1.320 \\
    Fe\,II\,$\lambda$4994 & 4994.74156 & 2.801 & -3.640 \\
    Fe\,II\,$\lambda$5019 & 5019.82798 & 2.886 & -1.220 \\
    Fe\,II\,$\lambda$5102 & 5102.27056 & 5.907 & -2.172 \\
    Fe\,II\,$\lambda$5138 & 5138.22476 & 2.840 & -4.290 \\
    Fe\,II\,$\lambda$5170 & 5170.45988 & 2.886 & -1.250 \\
    Fe\,II\,$\lambda$5236 & 5236.07213 & 3.216 & -2.230 \\
    Fe\,II\,$\lambda$5266 & 5266.26624 & 3.224 & -3.120 \\
    Fe\,II\,$\lambda$5273 & 5273.86566 & 5.948 & -2.029 \\
    Fe\,II\,$\lambda$5285 & 5285.56550 & 2.886 & -2.990 \\
    \bottomrule
    \end{tabular}
    \label{FeII_lines}
\end{table}

\section{Modelling the facular transit}\label{subsec:appendix_grid_model}

To model the solar surface, we use a $360^\circ$ longitude by $180^\circ$ latitude grid, such that the visible hemisphere comprises $180 \times 180$ pixels at a grid resolution of $1^\circ$. We model the transit of a circular patch of faculae with a radius of $15^\circ$ across the solar equator, from east to west. When seen at disk centre, this facular patch covers approximately 2.8\% of the visible disk\footnote{The chosen size is similar to that adopted for plage simulations in SOAP~2.0 \citep{Dumusque2014}.}.

To compute the disk-integrated flux, we sum the intensities of all pixels, weighted by their projected contribution and by the solid angle subtended by each surface pixel at 1~AU. For a pixel $(i,j)$, the weight is
\begin{equation}
w_{ij} = \mu_{ij}\,\Omega_{\text{pixel},ij},
\end{equation}
where $\mu_{ij}$ is the cosine of the viewing angle. The pixel solid angle is defined from the pixel surface area $dA_{ij}$ as
\begin{equation}
\Omega_{\text{pixel},ij} \equiv \frac{dA_{ij}}{\mathrm{AU}^2}.
\end{equation}
For a uniform longitude--latitude grid with angular spacing $\Delta_{lon} = \Delta_{lat} = \Delta = 1^\circ = \pi/180$, we approximate
\begin{equation}
dA_{ij} \approx R_\odot^2 \cos(l_{ij})\,\Delta^2,
\end{equation}
where $l_{ij}$ is the latitude of the pixels. The factor $\cos(l_{ij})$ accounts for the reduction in the pixel surface area towards the poles due to meridian convergence.

The total disk-integrated flux spectrum at time $t$ is then computed as

\begin{equation}\label{equ:disk-int}
\begin{aligned}
F_\mathrm{disk}(\lambda, t)
= \sum_{i,j} w_{ij} \cdot \bigl[
p_{ij}(t) \cdot I^{\mathrm{fac}}(\lambda, \mu_{ij}) + \\
\bigl(1 - p_{ij}(t)\bigr) \cdot I^{\mathrm{QS}}(\lambda, \mu_{ij})
\bigr].
\end{aligned}
\end{equation}

where $p_{ij}(t)$ is the facular filling factor at pixel $(i,j)$, equal to 1 if the centre of the pixel lies inside the prescribed circular facular region at time $t$, and 0 otherwise, such that the circular patch is realised as a pixelized mask on the discrete grid without fractional edge weighting. The intensity spectra $I^{\text{fac}}$ and $I^{\text{QS}}$ are assigned from precomputed QS and facular spectra (see Section \ref{subsec:spectral_input}). The intensity spectra themselves are time-independent, being averaged over the available snapshots (45 quiet and 40 facular snapshots); only the position of the facular patch changes with time. In the non-rotating case, both $I^{\text{QS}}$ and $I^{\text{fac}}$ are used without any additional rotation-induced Doppler shift. When solar rotation is enabled, each pixel’s local contribution is Doppler-shifted according to its projected rotational velocity:
\begin{equation}
\lambda'_{ij} = \lambda \left(1 + \frac{v_{ij}}{c} \right),
\end{equation}
where $v_{ij}$ is the LOS component of the solid-body rotational velocity at pixel $(i,j)$, and $c$ is the speed of light. The Doppler-shifted intensity is then evaluated by interpolating the static spectrum:
\begin{equation}\label{equ:Doppler_intensity}
I^{\text{rot}}(\lambda, \mu_{ij}) = I\left( \frac{\lambda}{1 + v_{ij}/c}, \mu_{ij} \right).
\end{equation}
In the rotating case ($F^{\text{rot}}_\mathrm{disk}(\lambda, t)$), we simply substitute the Doppler-shifted intensity from equation \ref{equ:Doppler_intensity} into equation \ref{equ:disk-int}.

\section{Flux response of the facular patch transit}\label{subsec:appendix_photometry}

The perturbations in the disk-integrated flux induced by the facular patch depend on its position on the solar disk during the transit and on the wavelength (continuum, left/right line wing, line core), as illustrated in Figure~\ref{fig:trailed_spectra}. These perturbations are shaped by solar rotation, the impact of magnetic field on convective velocities and the brightness contrast between faculae and QS.

In the rotating case, the facular transit produces a strongly wavelength-dependent and asymmetric flux difference signal across the spectral line (see Figures~\ref{fig:trailed_spectra}a and d) in agreement with the findings from \citet{Thompson2020}. As the facular patch moves across the disk, the line core shows a positive flux contribution relative to QS since faculae are brighter than QS in the line core \citep[see e.g.][for wavelength dependence of facular contrast]{Shapiroetal2016}, while the continuum exhibits a flux deficit near disk centre because faculae are darker than QS at disk centre (as discussed in Section~\ref{subsec:spectral_input}). A positive flux contribution is seen away from the disk centre owing to the increase in the facular contrast towards the limb. The wavelength--longitude pattern is clearly asymmetric about line centre and the central meridian, reflecting the combined effects of solar rotation and convective Doppler shifts within the facular atmosphere.

In the non-rotating case, the flux difference signal becomes symmetric in longitude while retaining its strong wavelength dependence (see Figures~\ref{fig:trailed_spectra}b and e). The facular region still produces a brightening in the line core and a continuum flux deficit near disk centre, showing that these features arise independently of rotation. The spectral signatures in this case are dominated by the CLV of contrast and convective effects within the faculae, with the strongest contrast occurring near central meridian passage. This case highlights that even without rotation, faculae generate non-negligible perturbations that vary across the line profile and contribute to the disk-integrated flux variability.

In the non-convective case, the flux difference signals represent the purely photometric contribution of the facular region (see Figures~\ref{fig:trailed_spectra}c and f). The flux difference becomes antisymmetric in wavelength and in longitude. The dominant feature is a continuum flux deficit at disk centre combined with enhanced contrast towards the limb, consistent with the centre-to-limb behaviour of facular brightness. The absence of wavelength-dependent asymmetries confirms that convective motions are essential for producing the line-profile distortions and Doppler-like signals seen in the rotating case, while photometric contrast alone cannot account for the full facular spectral signature.

\begin{figure}[ht!]
    \centering
    \includegraphics[width=0.8
        \linewidth]{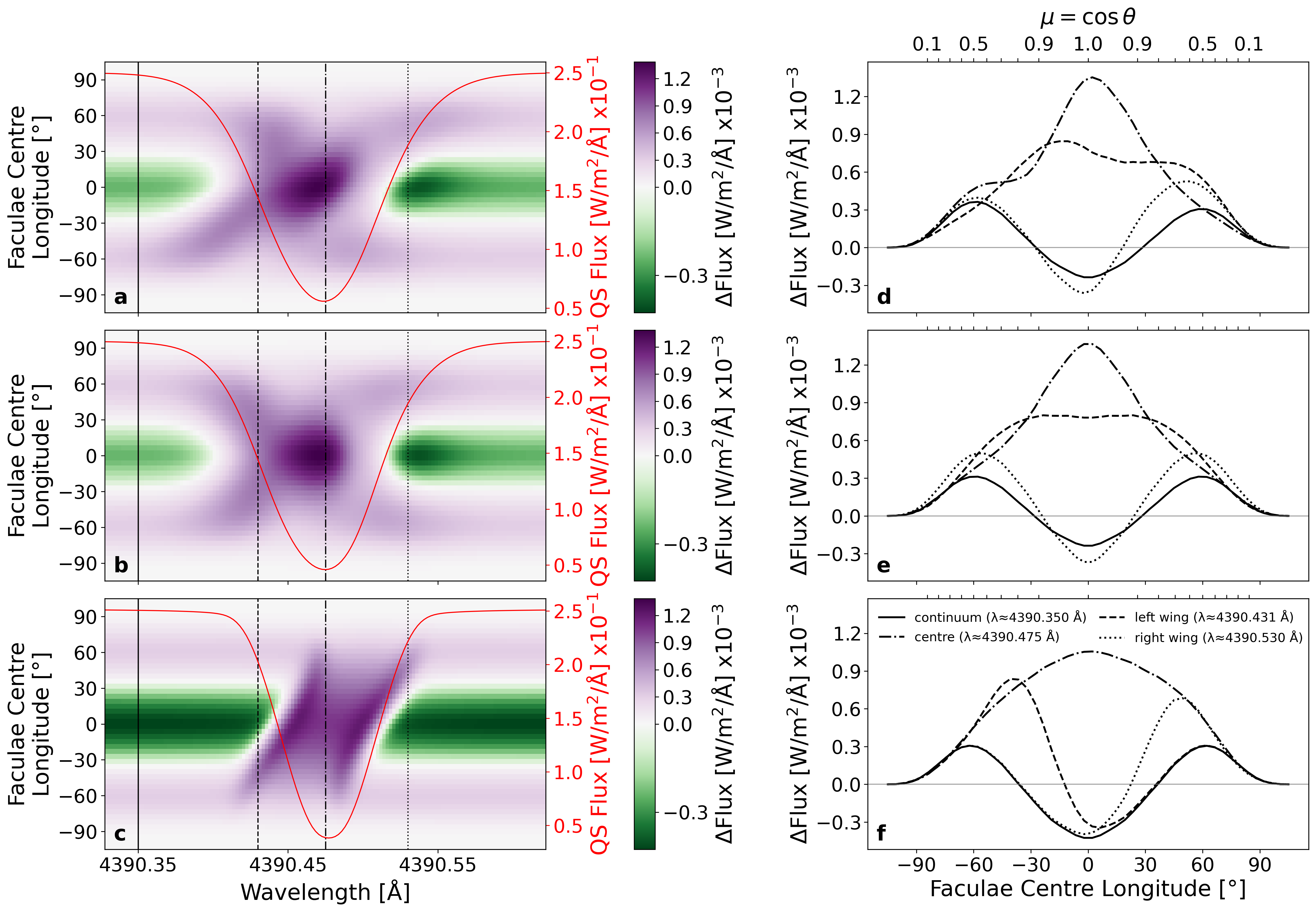}
    \caption{Panels (a)--(c): Time series of flux difference spectra ($\Delta F \equiv F_{\rm fac~transit}-F_{\rm QS}$) for the Fe\,I\,$\lambda$4390 line, covering a full disk transit of the facular patch. The flux differences are vertically stacked along the y-axis as a function of facular patch longitude. The y-axes are the same as the x-axes in Figure~\ref{fig:full_pic}. The red line in each panel represents the QS reference profile. Panel (a): Rotating case. Panel (b): Non-rotating case. Panel (c): Non-convective case (pure photometric signal from faculae). The colour scale encodes the differential flux, with purple indicating a positive flux difference (facular profile brighter than QS profile) and green regions indicating a flux deficit (faculae appearing dark in the continuum at disk centre, as introduced in Figure~\ref{fig:mu_spectra}). White marks equality between the QS and facular flux. Note the different scales between the purple and green colour as seen in the colour bars. Panels (d)--(f) show the flux profile of the respective vertical cuts through the colour plots on the left side. As a representative, we show 4 vertical cuts: through the continuum, the left wing of the line, the line centre, and the right wing (see legend in panel (f)).}
    \label{fig:trailed_spectra}
\end{figure}

\section{RV calculations}\label{subsec:appendix_rv_methods}

We use three complementary approaches to extract RVs from the continuum-normalised line profiles $f(\lambda, t)$. Note, we only present results for individual, isolated lines, where all measured shifts reflect the motion of a single line, unaffected by blending or neighbouring features.
The three estimators presented below were chosen as simple representatives of common RV extraction methods (flux-weighted COG shifts, Gaussian-fitted centroid RVs, and a linearised template matching projection). We did not optimise fine details of any of these algorithms. Our goal is to recover the overall physical RV signal associated with faculae, set by convective flows, their suppression in magnetised regions, facular brightening, and transit geometry.

\subsection{COG method}

For an isolated absorption line, the flux-weighted centroid of the absorption depth tracks the line’s COG. When a spectrum is Doppler-shifted by a small amount, the COG moves nearly linearly with the shift, providing a simple, model-independent centroid estimator. Compared to template matching, the COG can be more sensitive to local continuum placement and to profile asymmetries because all pixels in the window contribute with depth-based weights.

For each spectral line $k$ with laboratory wavelength $\lambda_{0,k}$, we consider a fixed window $[\lambda_{L,k},\,\lambda_{R,k}]$ around the line. Using $f(\lambda,t)$, the calculated spectrum at epoch $t$ during the facular transit, and $f_{\mathrm{QS}}(\lambda)$, the time-independent QS reference spectrum, we define the absorption depths as

\begin{equation}
a(\lambda,t) \equiv 1 - f(\lambda,t),
\qquad
a_{\mathrm{QS}}(\lambda) \equiv 1 - f_{\mathrm{QS}}(\lambda).
\end{equation}

\noindent The flux-weighted centroids of the line at epoch $t$ and for the QS reference are defined as

\begin{equation}
\begin{split}
\lambda_{\mathrm{COG},k}(t)
&= \frac{\displaystyle \int_{\lambda_{L,k}}^{\lambda_{R,k}} \lambda\, a(\lambda,t)\, d\lambda}
       {\displaystyle \int_{\lambda_{L,k}}^{\lambda_{R,k}}        a(\lambda,t)\, d\lambda}, \\
\lambda_{\mathrm{COG},\mathrm{QS},k}
&= \frac{\displaystyle \int_{\lambda_{L,k}}^{\lambda_{R,k}} \lambda\, a_{\mathrm{QS}}(\lambda)\, d\lambda}
       {\displaystyle \int_{\lambda_{L,k}}^{\lambda_{R,k}}        a_{\mathrm{QS}}(\lambda)\, d\lambda}.
\end{split}
\label{eq:cog-centroids}
\end{equation}

\noindent The Doppler shift relative to the QS reference is then estimated for each line as

\begin{equation}
\Delta v_k(t) \;\approx\; c\,
\frac{\lambda_{\mathrm{COG},k}(t) - \lambda_{\mathrm{COG},\mathrm{QS},k}}{\lambda_{0,k}},
\label{eq:cog-rv}
\end{equation}

\noindent which is the small-shift Doppler relation expressed with $\lambda_{0,k}$ in the denominator.

\subsection{Gaussian-fitting method}

We model the profile of the spectral line $k$ at an epoch $t$ (the in-transit spectrum with a local facular patch crossing the disk) and that from the QS reference with a Gaussian function given by

\begin{equation}
g(\lambda; A, b, \sigma, C) \;=\; A \exp\!\left[-\frac{(\lambda-b)^2}{2\sigma^2}\right] + C,
\end{equation}

\noindent fitted independently by least squares to obtain line centre estimates $\hat{b}_{\mathrm{fac},k}(t)$ and $\hat{b}_{\mathrm{QS},k}$. The constant term $C$ is included to absorb small residual continuum offsets (for ideally continuum-normalised profiles $C\!\approx\!1$); no slope term is fitted. The wavelength shift and RV for each line are then calculated using

\begin{align}
\Delta\lambda_k(t) &= \hat{b}_{\mathrm{fac},k}(t) - \hat{b}_{\mathrm{QS},k},\\
\Delta v_k(t) &\approx c\,\frac{\Delta\lambda_k(t)}{\hat{b}_{\mathrm{QS},k}},
\label{eq:nr-doppler}
\end{align}

\noindent where Eq.~\eqref{eq:nr-doppler} uses the non-relativistic Doppler approximation appropriate for small shifts. For synthetic, effectively infinite-S/N profiles the fits are unweighted.

Our approach is per-line and referenced to an infinite-S/N template, placing it within the general LBL framework. In contrast to the Gaussian fits to determine each line centre, most LBL implementations estimate per-line shifts by \emph{template matching in a small wavelength window} around each line \citep[as in][]{Dumusque2018}. We do not combine lines here; we only use the per-line RVs. The methodological difference (Gaussian fit vs. windowed template matching) does not affect the present analysis of perfectly isolated lines.

\subsection{Linearised template matching (projection onto the velocity derivative)}

In this method, we define the mean-subtracted flux difference

\begin{equation}
\begin{split}
\delta f(\lambda,t)
&= f(\lambda,t) - f_{\rm QS}(\lambda), \\
\tilde{\delta f}(\lambda,t)
&= \delta f(\lambda,t) - \langle \delta f(\cdot,t) \rangle_\lambda .
\end{split}
\end{equation}

\noindent where the angle brackets $\langle \cdot \rangle$ denote an average over wavelength. We project $\tilde{\delta f}$ onto the spectrum’s velocity derivative,

\begin{equation}
\begin{split}
f_v'(\lambda,t)
&\equiv \frac{\lambda}{c}\,\frac{\partial f}{\partial \lambda}(\lambda,t), \\
\tilde{f_v'}(\lambda,t)
&= f_v'(\lambda,t) - \langle f_v'(\cdot,t) \rangle_\lambda .
\end{split}
\end{equation}

\noindent and estimate the Doppler shift as

\begin{equation}
\Delta v(t) \;=\; -\,\frac{\displaystyle \int \tilde{\delta f}(\lambda,t)\,\tilde{f_v'}(\lambda,t)\,d\lambda}
{\displaystyle \int \big[\tilde{f_v'}(\lambda,t)\big]^2\,d\lambda} \, .
\label{eq:fdp-continuous}
\end{equation}

\noindent The minus sign follows from $f(\lambda,t)\approx f_{\rm QS}(\lambda-\delta\lambda)$ for a redshift, so $\delta f \simeq -\,f'_{\rm QS}\,\delta\lambda$ and the least squares solution gives a negative numerator for $\delta\lambda>0$.

This estimator is the standard first-order (small-shift) solution of template matching RVs obtained by projecting the residuals onto the template’s velocity derivative (see Eqs.~(1)--(6) in \citealt{Bouchy2001}), and the least squares template framework used in SERVAL (see Section~2 in \citealt{Zechmeister2018}, which explicitly discusses optimal weighting by local gradients).

Intuitively, a pure Doppler shift makes the spectrum change like its own slope; correlating the (mean-subtracted) difference spectrum with the (mean-subtracted) velocity derivative therefore picks out the shift while down-weighting flat pixels and continuum.

\subsection{Comparison of RV methods} 

\begin{figure}[ht]
    \centering
    \includegraphics[width=0.6
        \linewidth]{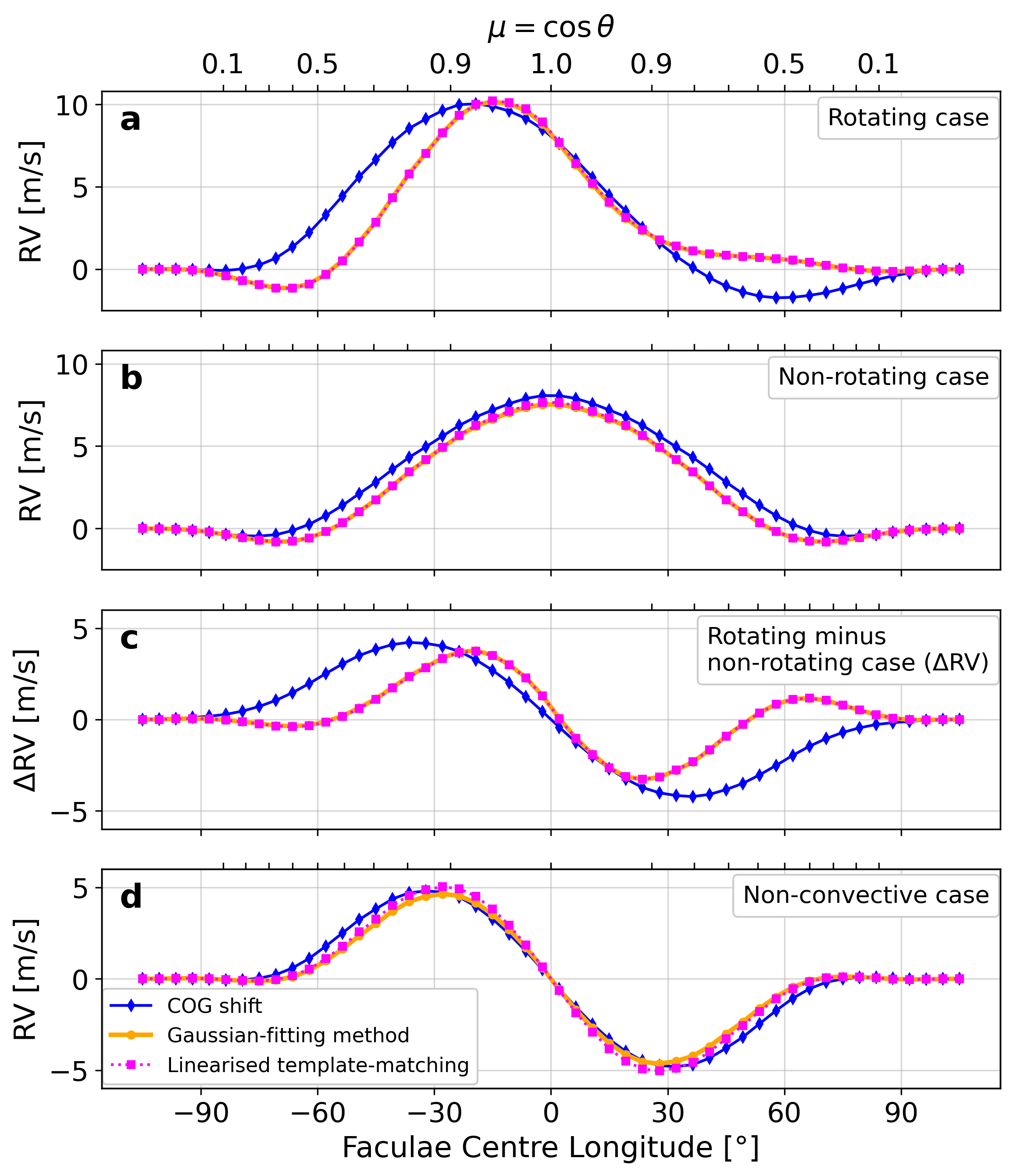}
    \caption{Facular transit RV profiles for the Fe\,I\,$\lambda$4390 line, calculated separately for the three RV methods introduced in Section~\ref{subsec:RV-methods}. Panel (a): Rotating case. Panel (b): Non-rotating case. Panel (c): Difference between rotating and non-rotating case. (d) Non-convective case (see Section~\ref{subsec:fac_rv}). The x-axes are the same as in Figure~\ref{fig:full_pic}. We note that in panels (a)--(c), the orange curve lies underneath the magenta curve.}
    \label{fig:rv_methods_FeI439}
\end{figure}

All three methods described above reproduce very similar facular transit RV profile amplitudes (Figure~\ref{fig:rv_methods_FeI439}), though they differ slightly in shape due to their intrinsic sensitivities to line asymmetries and broadening. Notably, the per-line centroid method (magenta) and linearised template matching method (orange) produce almost identical RV responses for isolated lines.
The COG method, on the other hand, differs intrinsically from the other two methods due to its reliance on the flux-weighted centroid. For simplicity and consistency, we average the results across all three methods. This approach helps to mitigate method-specific biases, as our lines are asymmetric, and different methods exhibit varying sensitivities to such asymmetries.

\section{Differential CB inhibition}\label{subsec:appendix_mu_inhibition}

The centre-to-limb behaviour of facular CB inhibition varies substantially from line to line and scales primarily with line strength. In particular, the $\mu$ location at which the signal changes sign, from CB inhibition (a relative redshift) near disk centre to CB enhancement (a relative blueshift) towards the limb, is highly line-dependent: weak (low-EW) lines tend to transition only at smaller $\mu$, whereas strong (high-EW) lines transition already at larger $\mu$ (Figure~\ref{fig:µ-dep_inhibition_multi-line}). In addition, the overall RV gradient across $\mu$ systematically steepens with increasing line strength.

Taken together, this structured $\mu$-dependence provides a physically motivated basis for the pronounced LBL spread in faculae-driven RV signatures discussed in Section~\ref{subsec:multi-line}.

\begin{figure}[t]
    \centering
    \includegraphics[width=0.6
        \linewidth]{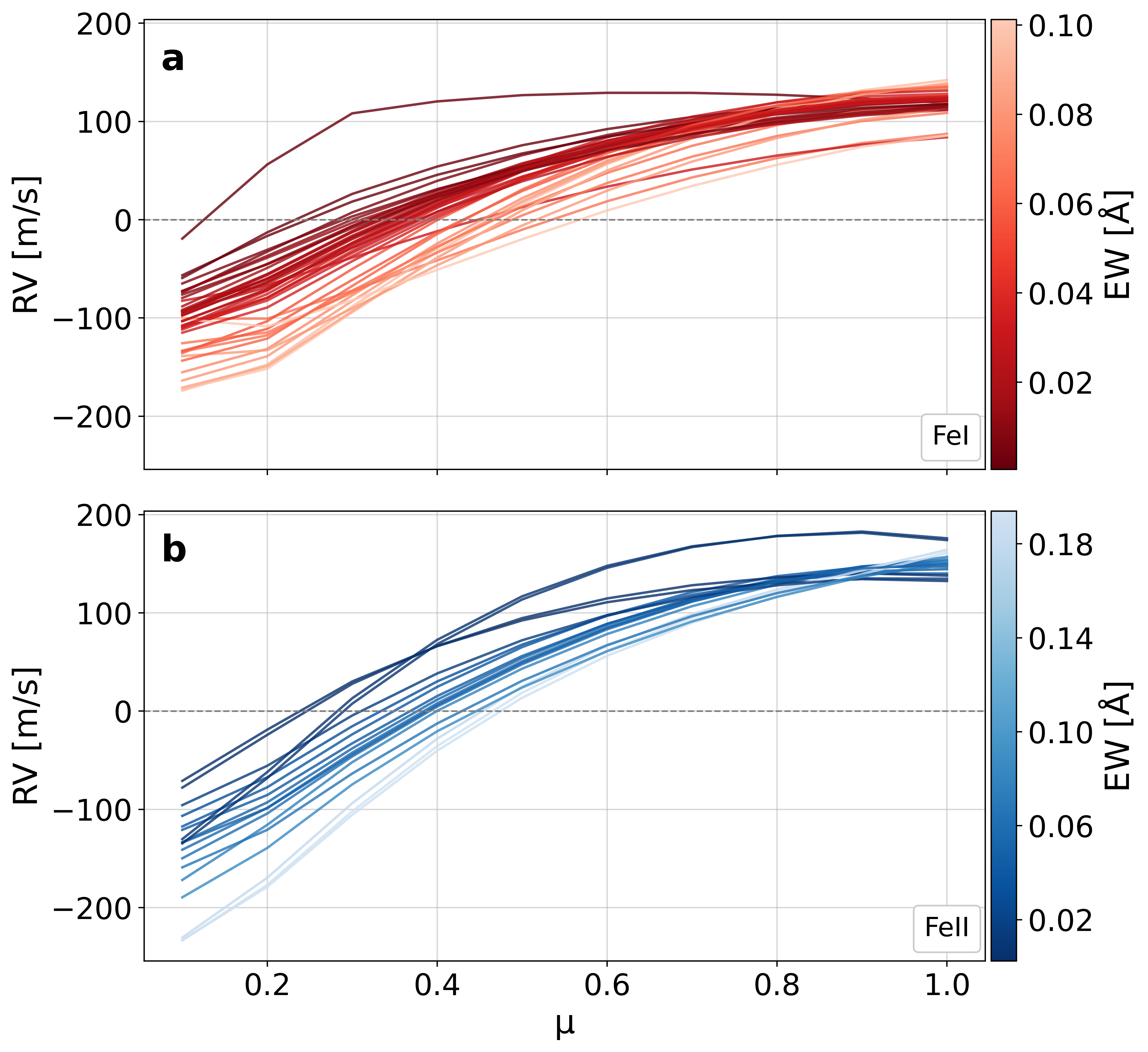}
     \caption{Centre-to-limb dependence of the RV signal due to facular suppression of CB for 60 isolated Fe\,I (panel a) and Fe\,II (panel b) lines. The RV signals are computed from the raw facular spectra, using the QS as reference. Line strength (disk centre EW) is encoded by colour saturation.}
    \label{fig:µ-dep_inhibition_multi-line}
\end{figure}

\section{Additional experiment scenarios: latitudes, filling factors, and rotation velocities}\label{subsec:appendix_add-scenarios}

To complement the baseline equatorial transit experiments, we tested how the disk-integrated faculae-induced RV signal changes under three key modifications: moving the facular transit to higher latitudes (changing viewing geometry and projection), changing the facular patch size, and increasing the solar rotation speed (changing the relative contribution of rotationally induced line-profile asymmetries). Unless stated otherwise, we use the Fe\,I\,$\lambda$4390 line and a facular patch angular radius of 15$^\circ$.

The RV amplitude decreases as the facular transit latitude increases, and the phase lag becomes smaller towards high latitudes (Figure~\ref{fig:rv_latitudes}a--b). This behaviour is consistent with the reduced rotational imprint at higher latitudes (smaller projected LOS rotation) and with the changing $\mu$-weighting sampled along the transit path across the visible disk. At latitudes of 60$^\circ$ and above, the facular transit samples only the limb where $\mu$ angles are smaller. Since at low $\mu$ angles, horizontal velocities dominate and lead to a blueshift (see Section~\ref{sec:Results}), the resulting RV signal is negative (blueshifted) throughout the transit.

The RV amplitude also scales approximately linearly with the facular patch size, i.e.\ with the facular filling factor. Small departures from strict linearity arise because projection effects and $\mu$-dependent weighting vary across the disk. Because we consider this linear behaviour with the filling factor to be trivial, we do not show an extra plot here.

Increasing rotation amplifies the rotationally induced line-profile asymmetries and drives the transit RV profile towards the characteristic `non-convective' (contrast-dominated) profile: a strong redshift on the approaching hemisphere and a strong blueshift on the receding hemisphere (Figure~\ref{fig:rv_latitudes}c; see the blue curve in Figure~\ref{fig:M-shape}b for the `non-convective' RV profile). As discussed in Section~\ref{subsec:fac_rv} and illustrated in Figure~\ref{fig:sketch_line}, this evolution reflects the increasing dominance of rotation-driven line skew induced by the facular contrast contribution, which progressively overwhelms the more subtle convective flow components.

\begin{figure}[ht]
    \centering
    \includegraphics[width=0.9
        \linewidth]{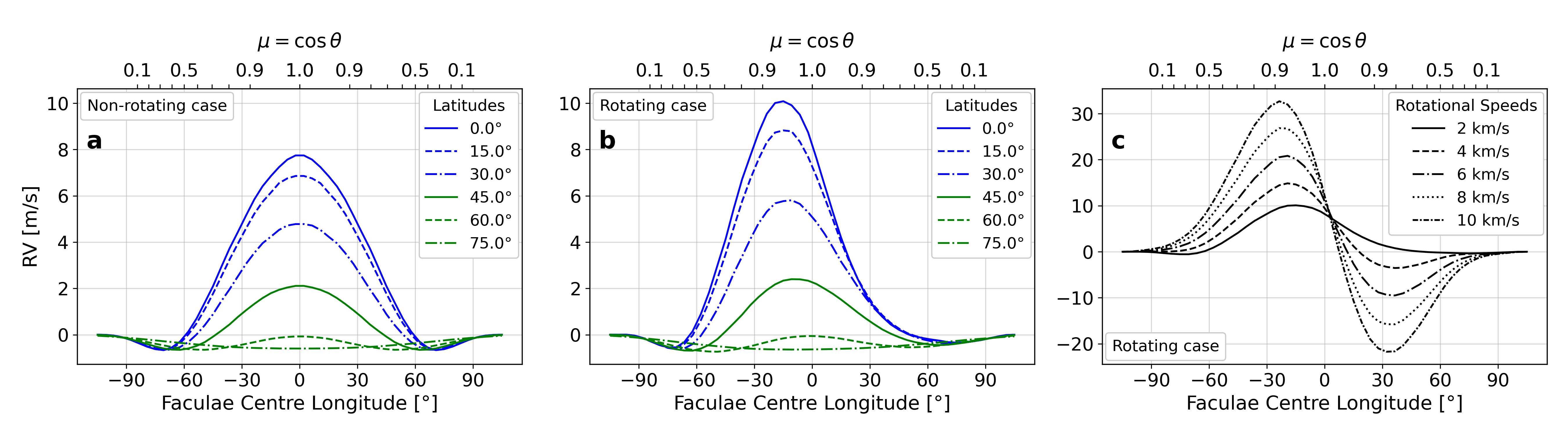}
    \caption{Facular transit RV profiles for the Fe\,I\,$\lambda$4390 line, in different model scenarios. The facular patch angular radius is 15$^\circ$. Panel (a): Non-rotating case: transits at latitudes 0$^\circ$--75$^\circ$ (legend). Panel (b): Rotating case at solar rotation ($v_{\rm eq}=2$~km/s): same set of latitudes. Panel (c): Equatorial transit (latitude 0$^\circ$) for different rotation speeds $v_{\rm eq}=$ 2, 4, 6, 8, and 10~km/s. The 2~km/s case in panel (c) (black solid curve) is the same as the equatorial case in panel (b) (blue solid curve). The x-axes are the same as in Figure~\ref{fig:full_pic}.}
    \label{fig:rv_latitudes}
\end{figure}

\section{Convective flows --- MURaM slices}\label{subsec:appendix_flow_physics}

In Section~\ref{subsec:inhibition} we use an idealised cartoon to build intuition for the $\mu$-dependent Doppler signal from granulation and faculae (Figure~\ref{fig:facula_sketch}). Here we provide representative Y--Z slices from a magnetic MURaM simulation (initialised with a 200\,G seed field) to emphasise that convective flows in and around facular magnetic flux concentrations are intrinsically complex and strongly height-dependent (Figures~\ref{fig:muram_6} and \ref{fig:muram_3}).

Even for comparatively isolated magnetic elements, the surrounding velocity field shows narrow flow channels, sharp shear layers, and rapid spatial sign changes on sub-granular scales, while the corrugated visible surface varies substantially across the magnetic and non-magnetic environment (black contour line in Figure~\ref{fig:muram_6}). In regions with clustered flux concentrations, the dynamics become more non-axisymmetric and heterogeneous, with interacting flow components and no one-to-one mapping between magnetic structure and local velocity component (Figure~\ref{fig:muram_3}).

These examples motivate treating facular Doppler signatures as an average over a broad distribution of local intensity--velocity configurations, and they support the interpretation that the $\mu$-dependence of the net RV can manifest itself as both reduced and, toward the limb, occasionally enhanced convective blueshifts, depending on geometry and magnetic structuring.

\begin{figure}[ht]
    \centering
    \includegraphics[width=0.75
        \linewidth]{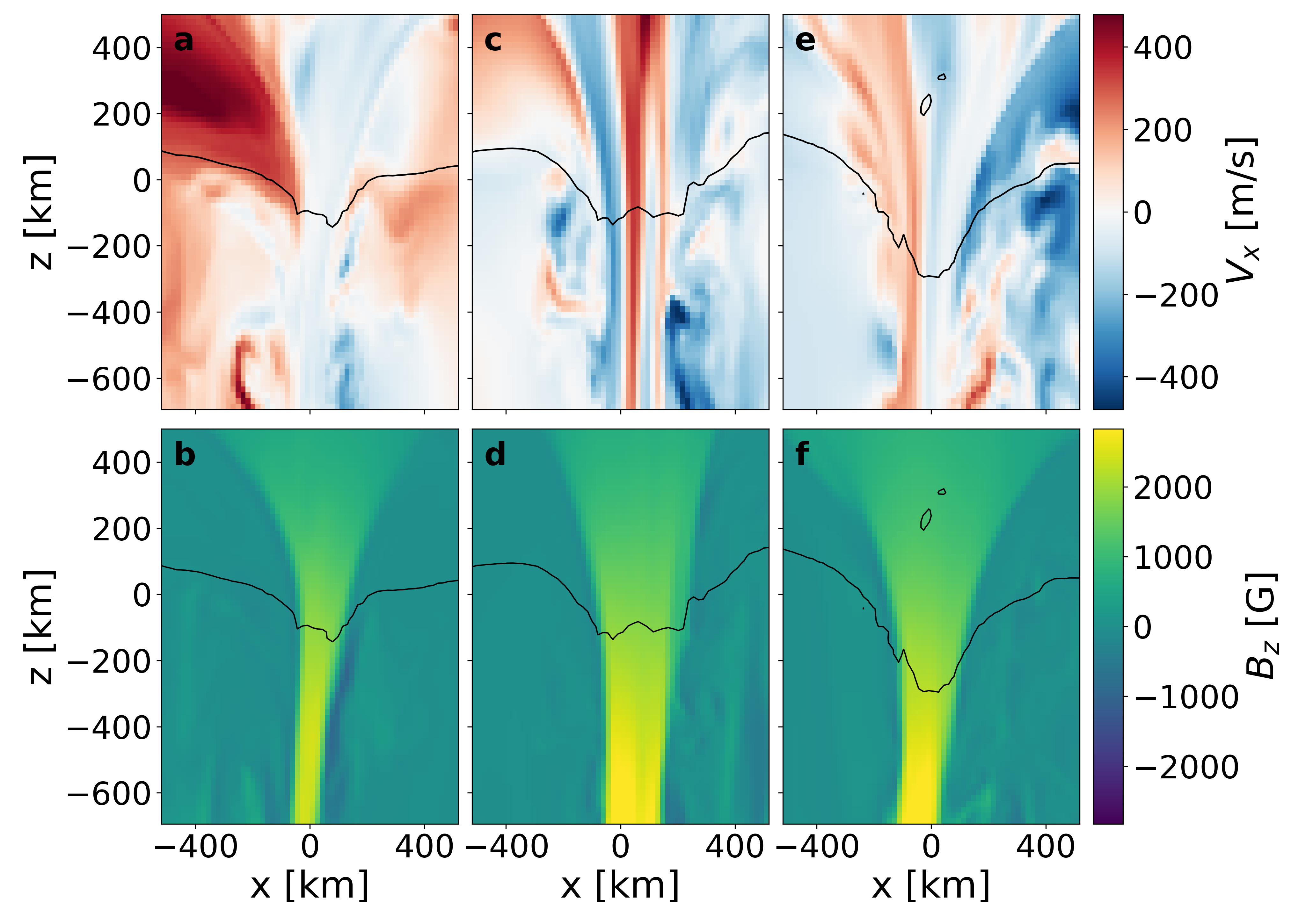}
    \caption{Representative slices through three comparatively isolated magnetic elements in a MURaM simulation cube (initialised with a 200\,G mean vertical seed field); each column corresponds to a different magnetic flux tube. The top row shows the horizontal velocity component in the $x$-direction ($V_x$), and the bottom row the vertical magnetic field component ($B_z$). The black contour marks the 6000\,K isotherm, used here as a proxy for the visible surface ($\tau \approx 1$), highlighting the local surface corrugation and the depression within the magnetic concentrations.}
    \label{fig:muram_6}
\end{figure}

\begin{figure}[ht]
    \centering
    \includegraphics[width=0.75
        \linewidth]{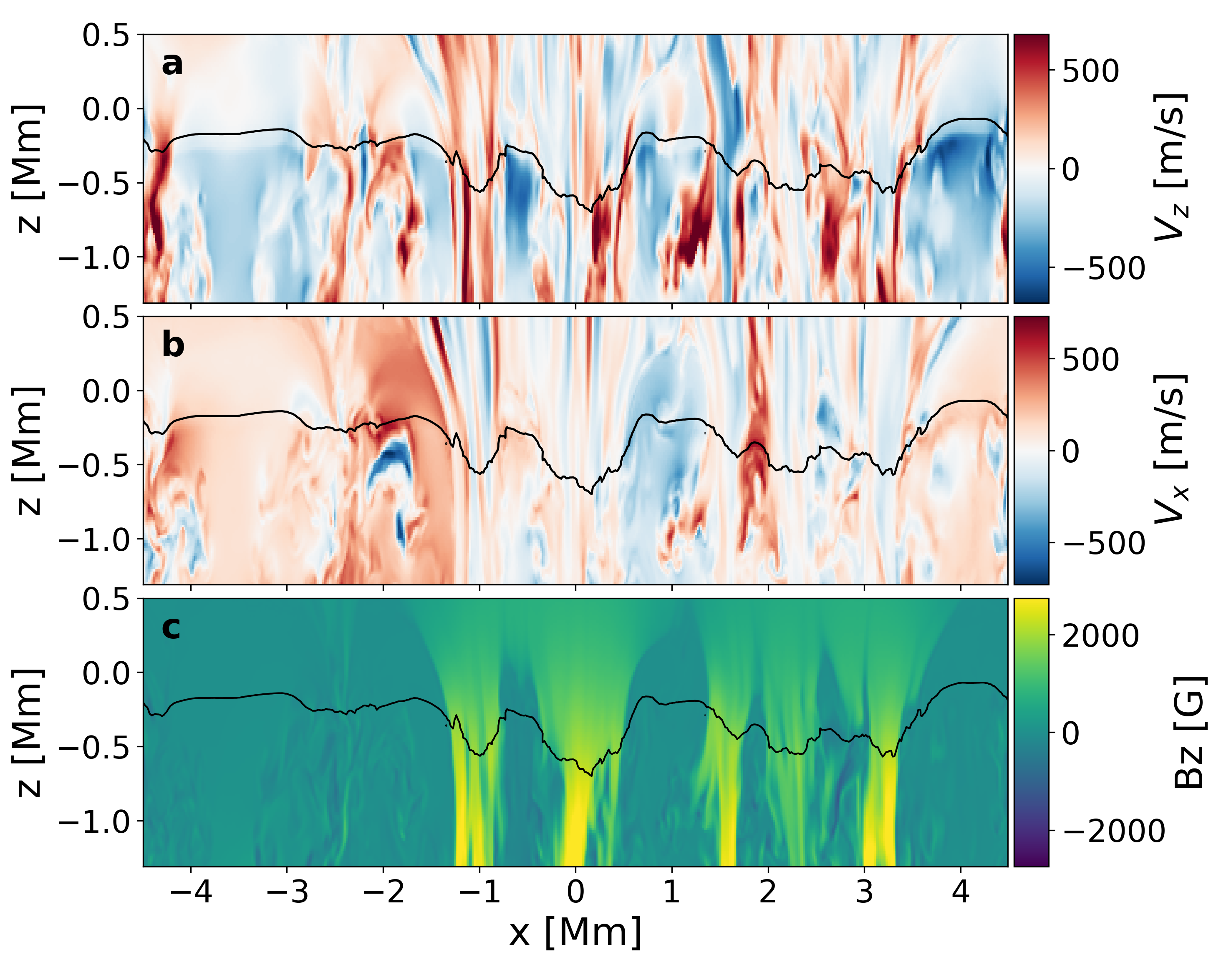}
    \caption{A slice through a region with multiple magnetic flux concentrations in a MURaM simulation cube (initialised with a 200\,G mean vertical seed field). Panels show (a) vertical velocity ($V_z$), (b) horizontal velocity ($V_x$), and (c) vertical magnetic field ($B_z$). The black contour marks the 6000\,K isotherm (proxy for $\tau \approx 1$), illustrating how clustered magnetic concentrations are embedded in a highly structured, non-axisymmetric, and height-dependent flow field.}
    \label{fig:muram_3}
\end{figure}

\end{document}